\documentclass{article}

\usepackage{arxiv}

\PassOptionsToPackage{numbers, compress}{natbib} 

\usepackage{etoolbox}
\apptocmd{\thebibliography}{\raggedright}{}{}
\usepackage[hyphens]{url}
\usepackage{breakurl}  
\usepackage{changepage}
\usepackage{float}
\usepackage{array}
\usepackage{natbib}
\usepackage{listings}
\usepackage{xcolor}
\usepackage{enumitem}   
\usepackage{threeparttable} 
\usepackage{setspace}    
\usepackage[utf8]{inputenc} % allow utf-8 input
\usepackage[T1]{fontenc}    % use 8-bit T1 fonts
\usepackage{hyperref}       % hyperlinks
\usepackage{url}            % simple URL typesetting
\usepackage{booktabs}       % professional-quality tables
\usepackage{amsfonts}       % blackboard math symbols
\usepackage{nicefrac}       % compact symbols for 1/2, etc.
\usepackage{microtype}      % microtypography
\usepackage{lipsum}
\usepackage{graphicx}
\graphicspath{ {./images/} }
\usepackage{parskip}
\setlist[enumerate]{itemsep=0pt, topsep=0pt}
\setlist[itemize]{itemsep=0pt, topsep=0pt}

\title{Enterprise Security Incident Analysis and Countermeasures Based on the T-Mobile Data Breach}        
\author{
	Zhuohan Cui \\
	The University of Sydney,\\ Camperdown NSW 2050, Australia\\
	\texttt{zcui0321@uni.sydney.edu.au}
	\And
	Zikun Song \\
	The University of Sydney,\\ Camperdown NSW 2050, Australia\\
	\texttt{zson0379@uni.sydney.edu.au}
}

\begin{document}
\maketitle
\begin{abstract}
This paper presents a comprehensive analysis of T-Mobile’s critical data breaches in 2021 and 2023, alongside a full-spectrum security audit targeting its systems, infrastructure, and publicly exposed endpoints. By combining case-based vulnerability assessments with active ethical hacking techniques—including Shodan reconnaissance, API misuse simulations, VNC brute-forcing, firmware reverse engineering, and web application scans—we uncover structural weaknesses persisting beyond the initial breach events. Building on these findings, we propose a multi-layered defensive strategy encompassing Zero Trust Architecture, granular role-based access control, network segmentation, firmware encryption using AES with integrity checks, and API rate limiting and token lifecycle control. Financial modelling demonstrates that a five-year investment yields less than 1.1\% of expected breach losses, validating the cost-effectiveness of proactive security measures. Our work bridges post-incident forensic analysis with hands-on security evaluation, providing an actionable blueprint for large-scale telecoms seeking operational resilience, regulatory compliance, and cross-domain threat readiness. 
\end{abstract}


\section{Introduction}
\label{sec:Introduction}

\subsection{Definition and Nature of Data Breaches}
\label{subsec:Definition}
A corporate data breach is a security incident in which unauthorised individuals gain access to sensitive, confidential, or protected data held by an organisation. This may involve the theft, exposure, or destruction of personally identifiable information (PII), financial records, intellectual property, or operational data. Data breaches can result from: External cyberattacks (e.g., phishing, ransomware, exploitation of vulnerabilities); Insider threats (malicious or accidental); System misconfigurations or human error; Third-party vendor compromise. 

\subsection{Impacts of Data Breach Events}
\label{subsec:Impacts}
The consequences of a data breach are multifaceted and severe: 
\begin{enumerate}[left=1em]

\item Financial Loss: Average global cost per breach reached 4.88 million US dollars in 2024, with healthcare and finance sectors experiencing even higher losses~\cite{ibm2024databreach}~\cite{secureframe2024databreach}. Costs include legal fees, regulatory fines, incident response, and customer compensation. 

\item Reputational Damage: Breaches erode consumer trust and brand equity, often leading to customer attrition and long-term market disadvantage~\cite{ibm2024databreach}~\cite{hbr2023cyberbreach}. 

\item Operational Disruption: Breaches may halt business operations, especially when ransomware or destructive malware is involved. 

\item Legal and Regulatory Consequences: Organisations may face penalties under laws such as GDPR\footnote{The General Data Protection Regulation (GDPR) is a comprehensive data privacy law enacted by the European Union in 2016 and enforced since May 25, 2018. It governs the collection, processing, and transfer of personal data, granting individuals greater control over their information and imposing strict compliance requirements on organisations worldwide that handle EU residents’ data.}, HIPAA\footnote{The Health Insurance Portability and Accountability Act (HIPAA) is a U.S. federal law enacted in 1996 that establishes national standards for protecting sensitive patient health information. It includes provisions for privacy, security, and breach notification, and applies to healthcare providers, insurers, and business associates that handle protected health information (PHI)\cite{cdcHIPAA1996}.}, or Australia’s Notifiable Data Breaches scheme. 

\item Identity Theft and Fraud: Exposed personal data can be exploited for financial fraud, phishing, and impersonation. 
\end{enumerate}

\subsection{Trends in Data Breaches in Recent Years}
\label{subsec:Trends}
Recent years have seen a marked escalation in both the frequency and scale of data breaches: 
\begin{enumerate}[left=1em]
\item Rising Incidence and Scale: In 2023 alone, over 353 million individuals were affected by publicly reported breaches~\cite{secureframe2024databreach}.  Notable breaches include Qantas (6 million customers), AT\&T (110 million), and Samsung (800 million records)~\cite{intellizence2024breaches}
.

\item Cloud and Third-Party Risks: Over 82\% of breaches involved cloud-stored data, and 98\% of organisations had at least one vendor breach~\cite{ibm2024databreach}~\cite{secureframe2024databreach}. 

\item Human Element and Credential Theft: 74\% of breaches involved human error or social engineering, and 86\% used stolen credentials~\cite{secureframe2024databreach}. 

\item Longer Detection Times: Average breach lifecycle: 204 days to detect, 73 days to contain, contributing to higher costs~\cite{secureframe2024databreach}.  

\item Sector-Specific Vulnerabilities: Healthcare, finance, and energy sectors remain top targets due to the sensitivity and value of their data~\cite{secureframe2024databreach}~\cite{statista2024breaches}.
\end{enumerate}

\subsection{T-Mobile 2021/2023 Data Breach Overview}
\label{subsec:Overview}
T-Mobile US, Inc. is the second-largest wireless carrier in the United States, serving over 130 million subscribers as of 2025~\cite{wikipediaTMobileUS}. Since 2018, T-Mobile has experienced intensive multiple data breaches, when looking at the results of these data breaches, it is not difficult to find that the impact of the August 2021 and January 2023 data breaches was particularly serious, and the amount of information leaked was huge, including not only account information, but also sensitive customer data, such as: names, addresses, dates of birth, Social Security numbers, phone numbers, emails, etc~\cite{firewallTimesTMobile}~\cite{wikipediaTMobileBreach}. This paper will focus on discussing the data breach incidents of T-Mobile in 2021 and 2023. 

In August 2021, the T-Mobile data breach event occurred when an attacker gained access to an unprotected GPRS gateway located in Washington. They then used a brute-force attack to achieve an SSH\footnote{Secure Shell (SSH) is a cryptographic network protocol that enables secure remote access, command execution, and file transfer over unsecured networks. It replaces older protocols like Telnet by providing encrypted communication between client and server systems.} login, despite the lack of controls to prevent multiple login attempts. After gaining access to the router, the attacker could move around the network due to a lack of network segmentation~\cite{wikipediaTMobileBreach}. T-Mobile identified 76 million current, former, and prospective customers in the US whose information might have been compromised in the data breach. This included: first and last names, addresses, dates of birth, Social Security numbers, and driver's licence numbers of 7.8 million current customers and around 40 million former and prospective customers; the names, dates of birth, and ID numbers of an additional 1.9 million former and prospective customers; and names, dates of birth, and often addresses of 6.1 million former and prospective customers. T-Mobile confirmed that no customer financial information, such as credit or debit card details, was exposed~\cite{fccTMobile2024}. 

In January 2023, T-Mobile reported to the U.S. Securities and Exchange Commission (SEC) that a "bad actor" exploited an unsecured application programming interface (API\footnote{An Application Programming Interface (API) is a set of protocols and tools that enable software components to communicate and interact with each other. APIs define how requests and responses are structured, allowing developers to integrate external services, automate workflows, and build scalable applications.}), resulting in the theft of personal data from approximately 37 million customer accounts. Stolen information included names, addresses, emails, phone numbers, birthdates, account details, and plan features—but did not include sensitive IDs or payment data. The breach activity began around November 25, 2022. T-Mobile became aware on January 5, 2023, and affected customers are being notified~\cite{secFilingTMobile2023}. Inadequate API security enabled unauthorised access to the database, akin to a previous data breach at Australian provider Optus in 2022, where hackers stole data from 10 million customers (a third of Australia's population)~\cite{krebsonsecurityTMobile2023}~\cite{wikipediaOptus2022}.  

In the following sections of this paper, we will analyse and compare the security issues in these two data breach incidents based on the above public information, and discuss the specific solutions to prevent such incidents from happening. After that, we will conduct security audits on some services and devices of T-Mobile that are currently exposed to the public, identify problems and provide solutions. 

\section{Vulnerability Assessment on 2021 \& 2023 Data Breach}
\label{sec:Vulnerability Assessment}

\subsection{Security Analysis of the 2021 Data Breach}
\label{subsec:Security Analysis of the 2021 Data Breach}
According to the news and disclosure in the T-Mobile data breach event in 2021, we discussed above, an attacker gained access to an unprotected GPRS gateway located in Washington, then used a brute-force attack to achieve SSH login; there were no controls to prevent multiple login attempts. After gaining access to the router, the attacker could move around the network due to a lack of network segmentation~\cite{wikipediaTMobileUS}. In this case, we can conclude that: 
\begin{enumerate}[left=1em]
	\item \textbf{Weak SSH Security Configuration:} The SSH security had been ignored or misconfigured (such as no restriction on attempts), so the attacker could perform a brute-force attack. Also, the SSH login password is weak, and the username is common, so the attacker could finally get the correct password through a brute-force attack.
	
	\item \textbf{Public Exposure of SSH Gateway:} The gateway exposed the SSH service to the attacker, which means there is no VPN\footnote{A Virtual Private Network (VPN) is a secure communication technology that encrypts internet traffic and masks the user's IP address, enabling private and anonymous browsing across public or untrusted networks. It is commonly used for data protection, remote access, and bypassing geographic restrictions.} to ensure the internal network is safe, and that only staff with privileges can access the internal network.
	
	\item \textbf{Poor Network Segmentation:} Inadequate segmentation allowed lateral movement throughout the network. Instead of minimal subnetting like \texttt{10.1.1.1/30} and \texttt{10.1.1.2/30}, engineers used overly broad configurations such as \texttt{10.0.0.0/8}, increasing the attack surface.
	
	\item \textbf{Lack of Multi-Factor Authentication (MFA):} Sensitive customer devices and information were accessible without MFA, enabling attackers to extract data directly.
	
	\item \textbf{Absence of Intrusion Detection/Prevention Systems (IDS/IPS):} No mechanisms existed to detect or block suspicious behaviour. This oversight allowed attackers to extract data from over 76 million users undetected.
	
	\item \textbf{Poor Data Protection Measures:} Customer data was not properly safeguarded within the internal network, making it vulnerable to unauthorised access and extraction.
	
	\item \textbf{Unprotected Routing Infrastructure:} Firewall misconfigurations left routers exposed to malicious traffic, failing to prevent internal data breaches once the attacker gained access.
	
	\item \textbf{Overreliance on Post-Incident Response:} T-Mobile's strategy leaned heavily on containment after incidents occurred, rather than proactive defence and real-time monitoring.
\end{enumerate}

\subsection{Security Analysis of the 2023 Data Breach}
\label{subsec:Security Analysis of the 2023 Data Breach}
In 2023, T-Mobile suffered a data breach when an attacker abused an API interface to continuously obtain user data for more than a month before it was discovered and stopped by security teams. 88\% of APIs had at least one vulnerability, and some vulnerabilities were dependent on each other~\cite{zhaoxia2017cloud}. The principal security vulnerabilities and possible vulnerabilities include: 
\begin{enumerate}[left=1em]
	\item \textbf{Inadequate API Access Control:} Low-permission attackers call API interfaces beyond their permissions and obtain a large amount of user data.
	
	\item \textbf{Imperfect Log and Monitoring System:} The data leakage lasted for more than a month, indicating that there were no relevant personnel to conduct a regular log audit. In addition, when the frequency, source IP, and access time were abnormal, there was no automatic alarm measure.
	
	\item \textbf{Absence of a Rate Limit:} The API lacks a rate limit (call frequency limit). The attacker calls the API several times within a specific period, and the system does not take any response measures.
	
	\item \textbf{Defective Token Authentication:} The absence of an authentication token allows low-permission attackers to access high-permission APIs directly. There is a token, but it expires and is not handled in time, allowing attackers to call a high-permission API a second time.
\end{enumerate}

\subsection{Comparative Analysis of 2021 and 2023 Vulnerabilities}
\label{subsec:Comparative Analysis}
\noindent\textit{A. Differences in Vulnerability Types}

In the T-Mobile data breach in 2021, hackers scanned an unprotected GPRS gateway server used for legacy mobile network connections, usually reserved for 2G/3G devices, which was open to the public because no strong authentication mechanism was activated. The attacker, John Erin Binns, used SSH brute force to gain access to a controlled host and move laterally, gaining access to more than 78 million stored customer data and losing more than \$500 million~\cite{wikipediaTMobileUS}.

In the T-Mobile data breach in 2023, hackers somehow discovered an unauthorised API. The API was open to the public, lacked the necessary authentication mechanism, and was not equipped with good security mechanisms. The attacker continued to call the API for 6 weeks before being discovered and fixed. During that time, more than 37 million sensitive information were leaked, and more than 10 million US dollars were lost~\cite{krebsonsecurityTMobile2023}.

\noindent\textit{B. Effectiveness and Shortcomings in Security }

\textbf{Effectiveness:} After the 2021 data breach, T-Mobile promptly invested \$150 million to consolidate its network security infrastructure, taking measures such as hardening patches, enhanced access controls, and network monitoring~\cite{fcc2024tmobile}. After the 2023 data breach, T-Mobile detected abnormal traffic in the sixth week and quickly fixed API-related vulnerabilities, protecting sensitive information from further leaks and preventing further losses~\cite{firewallTimesTMobile}.

\textbf{Shortcomings:} In 2021 data breaches, GPRS gateways as legacy systems should not be exposed to the public network and should be limited to the Intranet or controlled access. Being able to use SSH brute force means that there is no good login protection mechanism. The behaviour of lateral movement indicates the lack of a least privileged policy, behaviour auditing, etc. Data breaches in 2023 were a concern, with six weeks of multiple calls to the API. The API certification design was weak, and there was a lack of a reasonable call rate limit, a good log alarm mechanism, and regular log audits. 

\noindent\textit{C. Structural Issues in Risk Management Processes}

\textbf{Delayed Response:} In the 2021 incident, after obtaining the initial access rights, the attacker was able to freely move laterally in the company's internal network and obtain a large amount of sensitive data, indicating that the lack of real-time intrusion detection and response mechanism failed to block the attack behaviour in time at the initial stage of the intrusion. In the 2023 API breach, the attacker continued to call this unauthorised interface for six weeks before the abnormal traffic attracted attention, indicating that the enterprise's security operations centre failed to respond to persistent data access anomalies promptly. 

\textbf{Delayed Risk Identification:} In the 2021 incident, an outdated GPRS gateway server was exposed to the public network, indicating that T-Mobile failed to carry out effective asset inventory and classification management of legacy systems, and lacked the access path sorting and exposure surface assessment of critical services. In the 2023 incident, the interface was attacked due to a chronic lack of certification, a lack of access control, and a state of openness to the outside world. This shows that T-Mobile's life cycle management, registration, certification, and monitoring have a serious defect in the establishment of a robust API risk identification system. 

\section{Security Countermeasures in Response to T-Mobile's 2021 \& 2023 Data Breaches}
\label{sec:Security Countermeasures in Response}

\subsection{Targeted Improvement Security Measures}
\label{subsec:Targeted Improvement Security Measures}
\begin{enumerate}[left=1em]
	\item \textbf{Strengthen API Interface Security Management:} Implement a fine-grained access control mechanism for all API interfaces to ensure that access requests must pass strict authentication and permission verification before entering the system, preventing unauthorised users from obtaining sensitive data or operating system functions.
	
	\item \textbf{Build a Real-time Security Monitoring System:} Use the latest security monitoring technology to continuously track and dynamically analyse network traffic and user behaviour, identify potential risks in time, and improve the discovery and response speed of abnormal events.
	
	\item \textbf{Strengthen Employee Security Awareness Construction:} Led by the information security team, we regularly organise information security training and simulation drills for all employees to help them master the methods to identify common threats such as phishing, email fraud, and social engineering attacks, and enhance their overall protection ability.
\end{enumerate}

\subsection{Security Enhancement Plan}
\label{subsec:Security Enhancement Plan}
We conclude that the different plan types and security measures that T-Mobile should implement are listed in Table ~\ref{tab:tmobile-measures}. 

\begin{table}[htbp]
	\centering
	\caption{Security Measures T-Mobile Should Implement}
	\label{tab:tmobile-measures}
	\begin{tabular}{|>{\centering\arraybackslash}m{2.2cm}|>{\centering\arraybackslash}m{2.2cm}|>{\centering\arraybackslash}m{8.5cm}|}
		\hline
		\textbf{Plan Type} & \textbf{Duration} & \textbf{Detailed Measures} \\
		\hline
		Short-term & 0--6 months & Fix known vulnerabilities, deploy firewall devices at the network boundary, implement VPN to access the internal network through Internet, enhance API and network security, and establish basic monitoring and response mechanisms. \\
		\hline
		Mid-term & 6--12 months & Improve the zero-trust architecture, optimise access control policies, and enhance incident response capabilities. \\
		\hline
		Long-term & More than 1 year & Continuously assess and improve security strategies, implement periodic training on employees' safety awareness, adopt the latest security technologies and best practices, and ensure the continuous effectiveness of the security system. \\
		\hline
	\end{tabular}
\end{table}

\subsection{Specific Suggestions}
\label{subsec:Specific Suggestions}
\begin{enumerate}[left=1em]
	\item \textbf{Deploying Extended Detection and Response System (XDR):} Integrating multiple security monitoring sources, realising cross-platform threat identification and coordinated response, improving the discovery ability and processing efficiency of complex attack behaviours.
	
	\item \textbf{The Introduction of Security Operation Centre (SOC) Service:} To establish or entrust a third party to build a security operation centre, realise unified monitoring, centralised judgment, and rapid disposal of various security incidents, and improve the coordination of overall security protection.
	
	\item \textbf{Enhance Employees' Security Awareness \& Operational Standard Ability:} Conduct regular safety training and actual combat drills to help employees understand and apply the company's safety policy and master the ability to identify and respond to common risk scenarios.
	
	\item \textbf{Strengthen API Access Control:} When users log in, they get a token for additional authentication when calling the API, so that users can only call the API that meets their permissions.
	
	\item \textbf{Reduce the Time of Audit Log and Improve the Judgment of Abnormal Log During Audit:} Audit the log every half a month and adhere to investigating any anomaly during audit to reduce the error rate of abnormal logs.
	
	\item \textbf{Set an Appropriate Rate Limit:} When the same account calls the same API multiple times within a specific period, the account is temporarily blocked, and an abnormal account warning is displayed.
	
	\item \textbf{Token Authentication Optimisation:} The token generated when logging in to the account is destroyed immediately after the expiration of the validity period, reducing the possibility of the token being reused.
	
	\item \textbf{API Optimisation:} A clear version number must be specified for each API call to facilitate smooth upgrades.
\end{enumerate}

\subsection{Research into Best Security Practices}
\label{subsec:Research into Best Security Practices}

\noindent\textit{A. Implementing Zero Trust Architecture (ZTA)}

A zero-trust architecture means that anyone needs to authenticate to access system resources. Core practices include: 

\begin{enumerate}[left=1em]
	\item \textbf{Principle of Least Privilege:} Each user has only the permissions they need. If they can only read event A, then they will not have any permission other than reading event A.
	
	\item \textbf{Network Segmentation:} The company's internal network is refined into multiple isolated areas. When an area is accidentally breached, other areas are not allowed to access the area without permission, which increases the difficulty of lateral movement for attackers.
	
	\item \textbf{Continuous Authentication:} Although the user is logged in, they need to re-authenticate each time they access a resource or at regular intervals to ensure the legitimacy of their identity and behaviour.
	
	\item \textbf{Multi-Factor Authentication (MFA):} In the verification phase, such as when a user is logged in, the user's identity is further confirmed by text message, email, fingerprint, or authenticator in addition to account and password verification.
	
	\item \textbf{Data Encryption Protection:} Timely update and use the latest encryption strategy to encrypt the data in storage and transmission to prevent theft or tampering by attackers.
\end{enumerate}

\noindent\textit{B. Continuous Monitoring and Agile Incident Response}

Early detection and timely repair of vulnerabilities can effectively prevent security incidents such as data leakage: 

\begin{adjustwidth}{1em}{0pt}
\textbf{Continuous Monitoring Mechanism:} A real-time monitoring system is deployed in firewalls, gateways, API layers, and other places, which collects logs of all behaviours and is equipped with automatic analysis of abnormal logs and automatic alarm functions to improve the possibility of being detected in the early stage of security events.

\textbf{Agile Response Process:} Establish an efficient emergency response mechanism. When a security incident is detected, the relevant personnel will be notified, the vulnerability will be quickly identified and patched, and the system will be restored in time. In the whole process, all departments should ensure that the information they have is consistent and respond to the requirements of other departments quickly.
\end{adjustwidth}

\noindent\textit{C. Data Classification and Protection}

The data is classified, and different protection systems are adopted according to different levels: 

\begin{adjustwidth}{1em}{0pt}
\textbf{Precise Data Classification:} Usually, a five-level data classification model is used to classify data into five levels, such as public, internal, sensitive, confidential, and top secret. The API and related gateway used to obtain user data are usually classified as above.
	
\textbf{Supporting Processing System:} According to the level of data, it is equipped with the corresponding protection system, which mainly focuses on the rights of data storage, access, transmission, sharing and destruction. Security measures such as strong data encryption and strong identity authentication are usually set for confidential level or above.
\end{adjustwidth}

\noindent\textit{D. End-to-End Encryption (E2EE)}

The E2EE technology ensures that the data is encrypted from the source, and only the authorised receiver can decrypt and view the plaintext. Even if the hacker intercepts the data during transmission, the hacker can only see the ciphertext and cannot decrypt the plaintext: 

\begin{adjustwidth}{1em}{0pt}
\textbf{Use a Secure Communication Protocol:} TLS 1.3\footnote{Transport Layer Security (TLS) 1.3 is the latest version of the TLS protocol, standardized in RFC 8446, offering enhanced security and performance for encrypted communications. It eliminates outdated cryptographic algorithms, reduces handshake latency, and enforces forward secrecy by default, making it more resilient against modern cyber threats.} is the most used encryption protocol for data transmission, and any URL prefixed with \texttt{https://} uses this protocol to ensure data integrity.
	
\textbf{Comply with the Standard Specification:} Choose an industry-recognised encryption algorithm, such as AES, RSA, ECC, etc. The key should be kept secret and should not be held or obtained by anyone other than the administrator and authorised persons.
\end{adjustwidth}

\noindent\textit{E. Regular Third-Party Security Audits}

A third-party security audit is a security measure that the company regularly asks external security experts or institutions to check whether there are loopholes in the network system:

\begin{adjustwidth}{1em}{0pt}
\textbf{Clear Audit Scope:} Negotiate and confirm the evaluation object with the third party before the audit, such as identity systems, API interfaces, database authority management, gateways, and other critical areas.
	
\textbf{Make a Rectification Plan:} Following the audit findings, the third party shall provide timely feedback on existing vulnerabilities and offer reasonable repair suggestions to assist the company in addressing these vulnerabilities.
\end{adjustwidth}

\noindent\textit{F. Secure Design of Azure Resource APIs}
Azure Portal has achieved functionality, security, and auditability in API design. All resource management operations are completed through Azure Resource Manager (ARM) APIs\footnote{Azure Resource Manager (ARM) APIs are RESTful interfaces provided by Microsoft Azure that enable users to deploy, manage, and monitor cloud resources consistently and securely. They support role-based access control, template-based provisioning, and versioning to ensure scalable and auditable infrastructure operations.}, ensuring consistency across different clients such as the Portal, CLI, and SDK\footnote{A Software Development Kit (SDK) is a collection of tools, libraries, documentation, and code samples that enables developers to build applications for a specific platform, framework, or operating system. SDKs streamline development by providing standardised interfaces and essential resources.}. After users log into the Azure Portal, the system issues an Access Token based on the OAuth 2.0\footnote{OAuth 2.0 is an open-standard authorisation framework that enables applications to obtain limited access to user resources without exposing credentials. It uses access tokens issued by an authorisation server to securely delegate permissions across web, mobile, and cloud platforms.} protocol. It performs identity verification and permission checks for each API request using role-based access control (RBAC)\footnote{Role-Based Access Control (RBAC) is a security model that restricts system access based on a user's assigned role, streamlining permission management and enforcing the principle of least privilege. By grouping users according to job responsibilities, RBAC simplifies authorisation, enhances compliance, and reduces the risk of unauthorised access across applications and infrastructure.}. Identity verification includes username, password, token, certificate, single sign-on (SSO)\footnote{Single Sign-On (SSO) is an authentication mechanism that allows users to access multiple independent applications or systems using a single set of login credentials. By streamlining identity verification, SSO enhances user convenience, reduces password fatigue, and improves security through centralised access control.}, etc. Authorisation is conducted using OAuth 2.0 to restrict the operations that client applications can perform on behalf of users, ensuring that requesters can only access authorised resources~\cite{azureResourceManagerOverview}. 

\begin{figure}[htbp]
	\centering
	\includegraphics[width=0.8\linewidth]{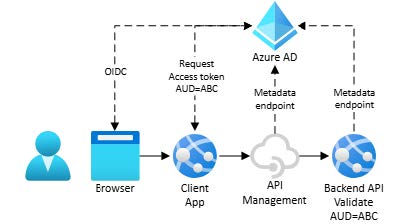}
	\caption{\href{https://learn.microsoft.com/en-us/azure/api-management/authentication-authorization-overview}{API Management Secure Access Architecture}}
	\label{fig:system-arch}
\end{figure}

In the design of Azure's API, version control (API Versioning) mechanisms are also employed. When the API is called, a specific version number must be specified to support backward compatibility and ensure smooth upgrades. Additionally, Azure implements rate-limiting policies to prevent misuse of API interfaces and uses HTTPS encryption to ensure the confidentiality and integrity of data transmission. Logs record detailed information about API calls, facilitating regular security audits and anomaly detection by security personnel~\cite{azureAPICACert}. The API management secure access architecture is shown in Figure ~\ref{fig:system-arch}.

\noindent\textit{G. Peer Industry Case Studies: Verizon and AT\&T}

\noindent\textbullet\ \textbf{Verizon – Proactive Threat Intelligence and Risk Management}
 
The core of Verizon's security strategy lies in the deep integration of continuous threat monitoring and data analysis capabilities. Its famous Data Breach Investigation Report (DBIR)\footnote{The Data Breach Investigations Report (DBIR) is an annual publication by Verizon that analyses thousands of real-world cybersecurity incidents and confirmed data breaches. It provides statistical insights into attack patterns, threat actors, and vulnerabilities across industries, serving as a foundational resource for improving organizational security strategies.} is not only a critical security research result in the world, but also an essential basis for enterprise internal security optimisation~\cite{verizon2025dbir}.

\hspace{1em}\textbf{\textit{Key Practices:}}

\hangindent=1em
\hangafter=0
Centralised incident management using a Security Operations Centre (SOC). \\
Data classification and access controls aligned with compliance frameworks (e.g., HIPAA, GDPR). \\
Regular penetration testing and real-time behavioural analytics. \\
Verizon's strong emphasis on continuous improvement and visibility into threat vectors has helped it reduce breach impacts and support customer trust~\cite{verizon2025dbir}.

\noindent\textbullet \textbf{AT\&T – Comprehensive Managed Security Services and Zero Trust Initiatives}

The core of AT\&T's security strategy is to integrate security capabilities in the form of "service and zero trust" and form a unified system in internal and external services. Its security solution emphasises the full synergy of cloud-native architecture, intelligent detection, and access boundary control~\cite{attZeroTrustBrief}~\cite{attPaloAltoAI}.

\hspace{1em}\textbf{\textit{Key Practices:}}

\hangindent=1em
\hangafter=0
Implement Zero Trust Network Access (ZTNA) throughout the organisation to ensure that only authorised identities can access system resources. \\
Deployed an Extended Detection and Response (XDR) platform to enhance cross-device and cross-application threat linkage processing capabilities. \\
Applying artificial intelligence and machine learning technologies to improve the automation level of user behaviour analysis and anomaly identification~\cite{attZeroTrustBrief}~\cite{attPaloAltoAI}. 

\section{Comprehensive Infrastructure and Security Enhancement Strategy for T-Mobile}
\label{sec:Comprehensive Infrastructure and Security Enhancement Strategy}
Based on the above analysis of T-Mobile’s data breach issues, we propose a series of solutions to address the security problems which mentioned above. 

\subsection{Network Topology Redesign and Segmentation}
\label{subsec:Network Topology Redesign and Segmentation}
\begin{figure}[htbp]
	\centering
	\includegraphics[width=0.8\linewidth]{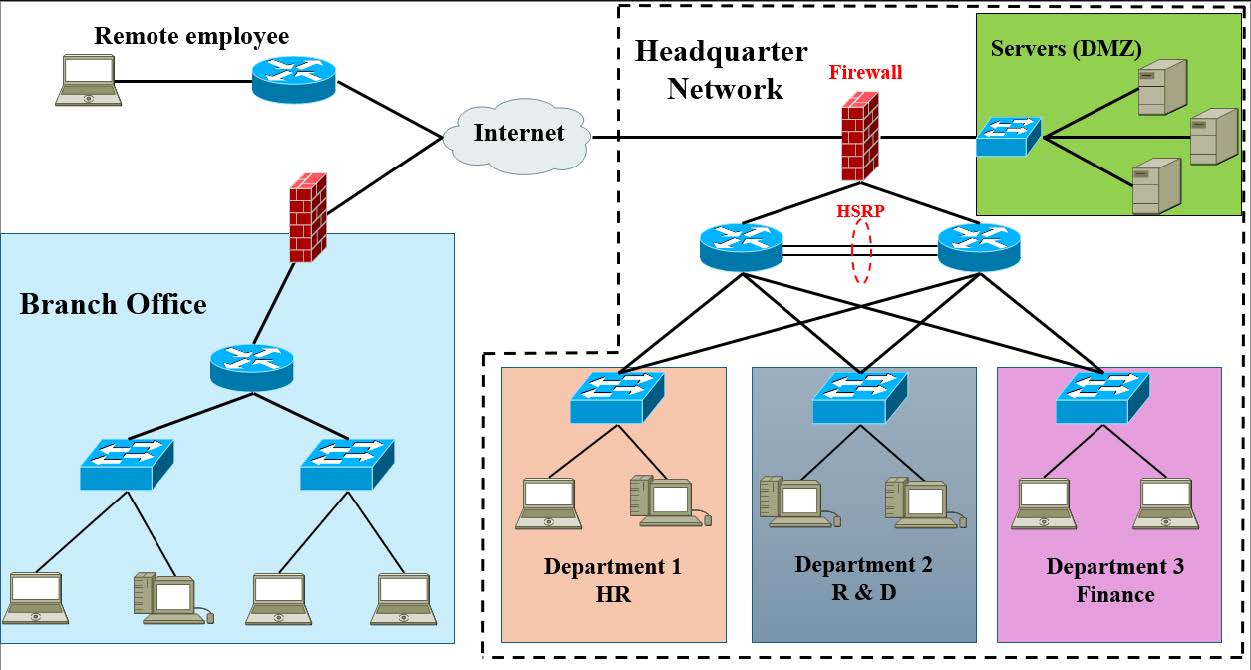}
	\caption{Network Design Logic Topology}
	\label{fig:network-logic-topology}
\end{figure}

Based on the analysis of the T-Mobile network before the data breach in 2021, we have redesigned a new brief logic network topology for the internal network (as shown in Figure ~\ref{fig:network-logic-topology}), and we have ignored some detailed designs, but focused on the significant problem of its network architecture. Also, we recommend that T-Mobile use the unified network devices manufacturer Cisco as their provider. 

\begin{figure}[htbp]
	\centering
	\includegraphics[width=0.8\linewidth]{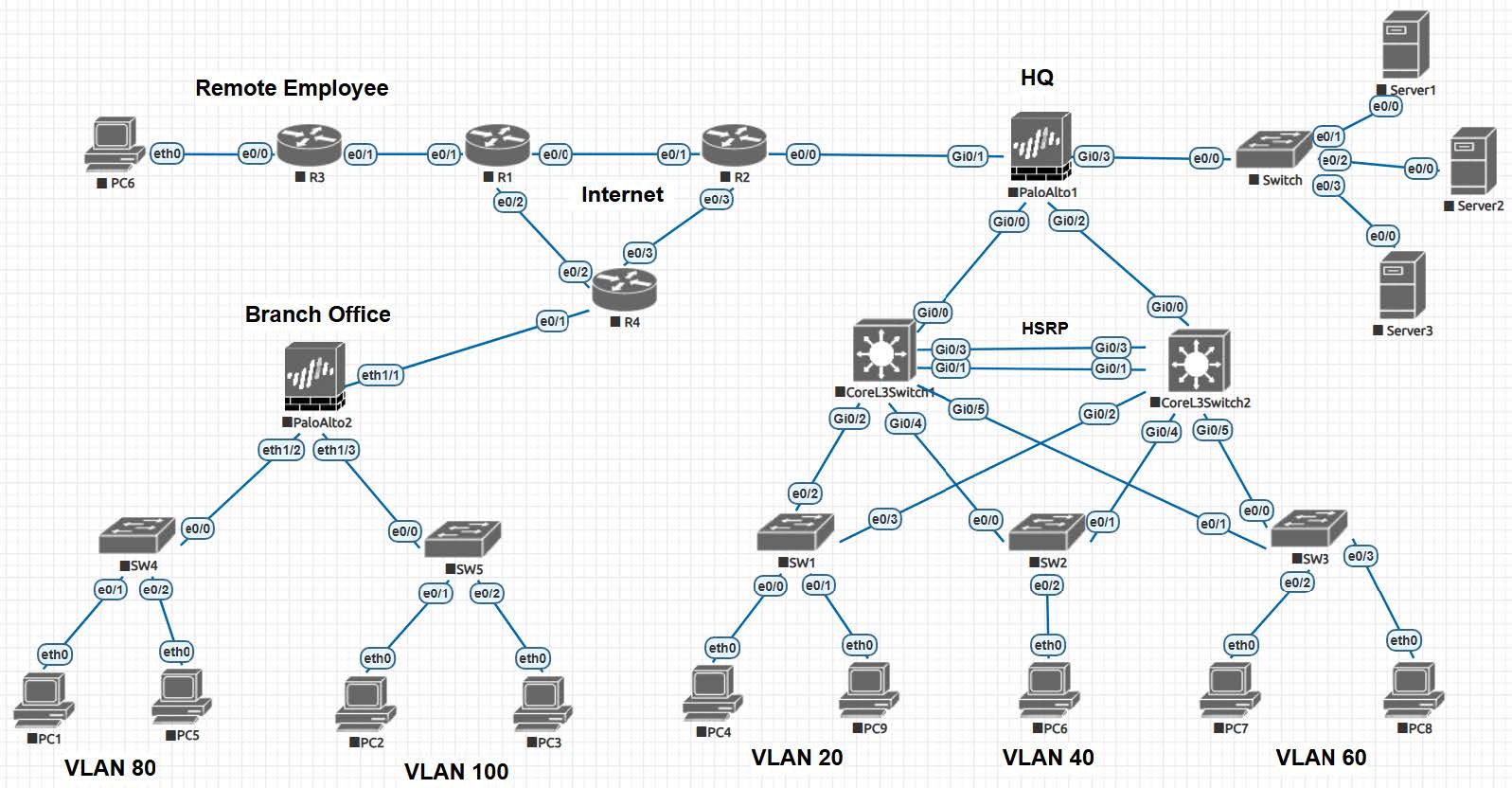}
	\caption{Network Design Topology in EVE}
	\label{fig:network-topology}
\end{figure}

In the above topology, every line means there is an Ethernet connection between the devices, and all devices belonging to the company are connected beyond the firewall and protected by the firewall. Then, if remote employees or branch offices want to access HQ's devices, they must use IPSec\footnote{Internet Protocol Security (IPSec) is a suite of protocols designed to secure IP communications by authenticating and encrypting each IP packet in a data stream. It operates at the network layer, supporting features like data integrity, origin authentication, and confidentiality, and is commonly used in VPNs to establish secure tunnels over public networks.} over VPN to connect to HQ's internal network, and each of the employees has a unique username and password for VPN. Also, every time you connect to the VPN, you need MFA to verify your identity again. Finally, we use Hot Standby Router Protocol (HSRP)\footnote{Hot Standby Router Protocol (HSRP) is a Cisco proprietary first-hop redundancy protocol that enables multiple routers to work together to present a single virtual default gateway to hosts on a LAN. It ensures high availability by designating one router as active and another as standby, automatically transferring routing responsibilities if the active router fails.} on core routers or layer 3 switches to ensure the availability of the internal network. We use HSRP to provide redundancy for the cluster of core routers. To briefly show the details of the correct network segmentation of the company's internal network, we will use EVE-NG\footnote{EVE-NG (Emulated Virtual Environment – Next Generation) is a clientless, multi-vendor network emulation platform that enables users to design, test, and troubleshoot complex network topologies using real virtual machines. It supports integration with various network and security devices, offers a web-based interface, and is widely used for certification training, proof-of-concept labs, and enterprise simulations.}, a platform widely used for simulating Cisco-based network environments. The topology is show in Figure \ref{fig:network-topology}

In both headquarters and branch office, we use Palo Alto firewall\footnote{Palo Alto Networks firewall is a next-generation security appliance that combines advanced threat prevention, application-aware traffic control, and deep packet inspection. Powered by PAN-OS, it leverages technologies like App-ID, User-ID, and Content-ID to provide granular visibility and enforcement across physical, virtual, and cloud environments.} as boundary device, and in each firewall should have at least three zones: trust, untrust and DMZ\footnote{A Demilitarized Zone (DMZ) is a network segment that acts as a buffer between an organization's internal network and external networks such as the internet. It hosts public-facing services like web, email, and DNS servers, allowing controlled access while isolating sensitive internal systems behind firewalls to enhance security and reduce exposure to cyber threats.}, trust zone includes interface Gi0/0 and Gi0/2 which is the highest level of security, untrust zone is interface Gi0/1 which is outside interface has the minimum level of security, DMZ include interface Gi0/3 which is the server zone has the medium level of security. In the internal network of each department should be divided by VLAN\footnote{A Virtual Local Area Network (VLAN) is a logical partitioning of a physical network that groups devices into isolated broadcast domains at the data link layer (OSI Layer 2). VLANs enhance network performance, security, and manageability by allowing administrators to segment traffic based on function, department, or application—regardless of physical location.} and belong to a different subnet, then, depending on the requirements to decide they could access other subnets, servers, or connect to the Internet, if necessary, implementing isolation, and finally, authenticate before doing any operation in the internal network. In this part, the company should implement the least privilege principle and ZTA. And configure the BPDU Guard\footnote{Bridge Protocol Data Unit (BPDU) Guard is a network security feature that protects Spanning Tree Protocol (STP) topologies by disabling ports that unexpectedly receive BPDUs. Typically enabled on access ports with PortFast, it prevents unauthorized switches or misconfigurations from influencing STP calculations, thereby mitigating risks like Layer 2 loops and topology manipulation.} in every interface on the switch that does not connect to the switch device, in case of a BPDU attack from inside. For availability and security, the company should configure LACP between the interfaces which connect two devices, and the highest level of STP (Spanning Tree Protocol)\footnote{Spanning Tree Protocol (STP) is a Layer 2 network protocol standardized as IEEE 802.1D that prevents switching loops by creating a loop-free logical topology in Ethernet networks. It dynamically disables redundant paths and designates a single active route between devices, ensuring reliable data forwarding and network stability.} of half VLANs and HSRP in core layer 3 switch and the second highest level of the other half of VLANs and HSRP in the other core layer 3 switch which ensure when one device is down, the other will take over the job. For instance, in the figure above, CoreL3Switch1 should configure as the VLAN 20 and 40 subnets' default gateway, and CoreL3Switch2 configure as VLAN 60 subnet's default gateway, if CoreL3Switch1 malfunctions, CoreL3Switch2 will take over its role as the gateway for VLAN 20 and 40 subnets, however, once CoreL3Switch1 returns to normal, it will reclaim its role as the gateway for VLAN 20 and 40 subnets. The same applies if SW2 malfunctions. This ensures the availability of the entire network. Then, about network segmentation, we present the following table to show the subnet division of the headquarters content as an example. The company should follow this division principle: For instance, VLAN 20 has approximately 50 people working, VLAN 40 has 230 people working, and VLAN 60 has 312 people working. Considering both redundancy and scalability, as well as not allocating excessive IP addresses, the subnet division list based on VLAN and interface IP address division for HQ is as follows: 

\begin{table}[htbp]
	\centering
	\caption{Subnet Division Based on VLAN}
	\label{tab:vlan-config}
	\begin{tabular}{|>{\centering\arraybackslash}m{3cm}|
			>{\centering\arraybackslash}m{4cm}|
			>{\centering\arraybackslash}m{4cm}|
			>{\centering\arraybackslash}m{3cm}|}
		\hline
		\textbf{VLAN Number} & \textbf{Subnet} & \textbf{Gateway IP Address} & \textbf{Distributable IP Capacity} \\
		\hline
		20 & 192.168.20.0/26 & 192.168.20.62 & 62 \\
		\hline
		40 & 192.168.40.0/24 & 192.168.40.254 & 254 \\
		\hline
		60 & 192.168.60.0/23 & 192.168.61.254 & 510 \\
		\hline
	\end{tabular}
\end{table}

\begin{table}[htbp]
	\centering
	\caption{IP Division Based on Interface}
	\label{tab:device-ip}
	\begin{tabular}{|>{\centering\arraybackslash}m{3.5cm}|
			>{\centering\arraybackslash}m{3cm}|
			>{\centering\arraybackslash}m{6cm}|}
		\hline
		\textbf{Device Name} & \textbf{Interface} & \textbf{IP Address/CIDR} \\
		\hline
		PaloAlto1 & Gi0/0 & 192.168.1.1/30 \\
		\hline
		PaloAlto1 & Gi0/2 & 192.168.2.1/30 \\
		\hline
		PaloAlto1 & Gi0/3 & 192.168.3.1/28 \\
		\hline
		CoreL3Switch1 & Gi0/0 & 192.168.1.2/30 \\
		\hline
		CoreL3Switch1 & VLAN-IF 20 & 192.168.20.62/26 \\
		\hline
		CoreL3Switch1 & VLAN-IF 40 & 192.168.40.254/24 \\
		\hline
		CoreL3Switch1 & VLAN-IF 60 & 192.168.61.253/23 \\
		\hline
		CoreL3Switch2 & Gi0/0 & 192.168.2.2/30 \\
		\hline
		CoreL3Switch2 & VLAN-IF 20 & 192.168.20.61/26 \\
		\hline
		CoreL3Switch2 & VLAN-IF 40 & 192.168.40.253/24 \\
		\hline
		CoreL3Switch2 & VLAN-IF 60 & 192.168.61.254/23 \\
		\hline
	\end{tabular}
\end{table}

\subsection{Enhancing API Security and System Integrity}
\label{subsec:Enhancing API Security and System Integrity}
Based on the analysis of the 2023 data breach on the T-Mobile event, we recommend the following cloud solutions (AWS\footnote{Amazon Web Services (AWS) is a comprehensive cloud computing platform provided by Amazon, offering scalable infrastructure and services such as computing, storage, networking, databases, and machine learning. It enables organizations to build, deploy, and manage applications securely and efficiently across global data centres using a pay-as-you-go model\cite{aws_homepage}.}): 

\noindent\textit{A. Strengthen API Access Control}

\begin{figure}[htbp]
	\begin{minipage}[t]{0.48\textwidth}
	\centering
	\includegraphics[width=1\linewidth]{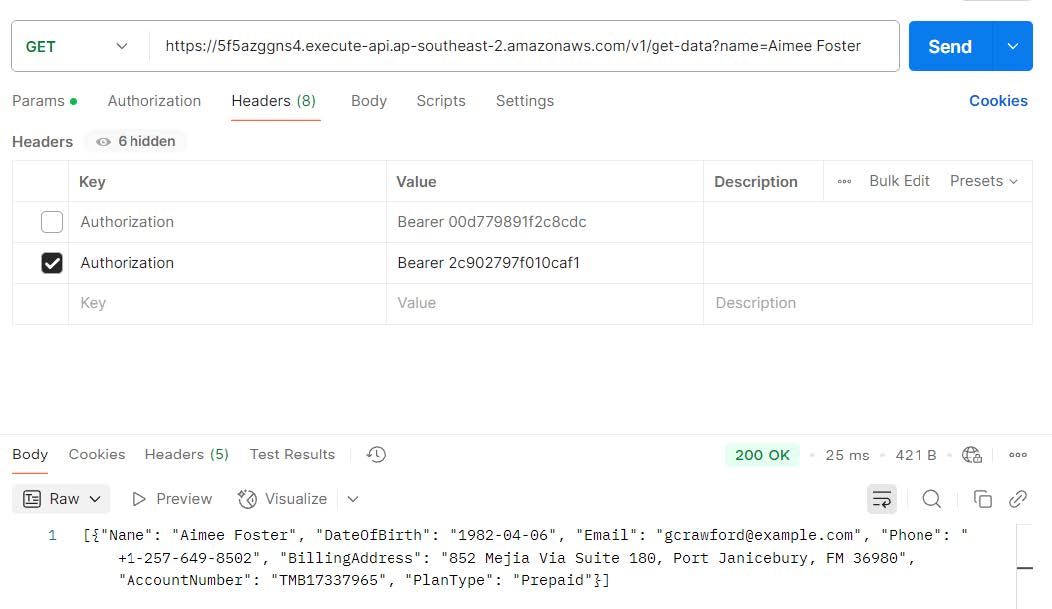}
	\caption{User Permission Test Result}
	\label{fig:user-test-result}
	\end{minipage}
	\hfill
	\begin{minipage}[t]{0.48\textwidth}
	\centering
	\includegraphics[width=0.8\linewidth]{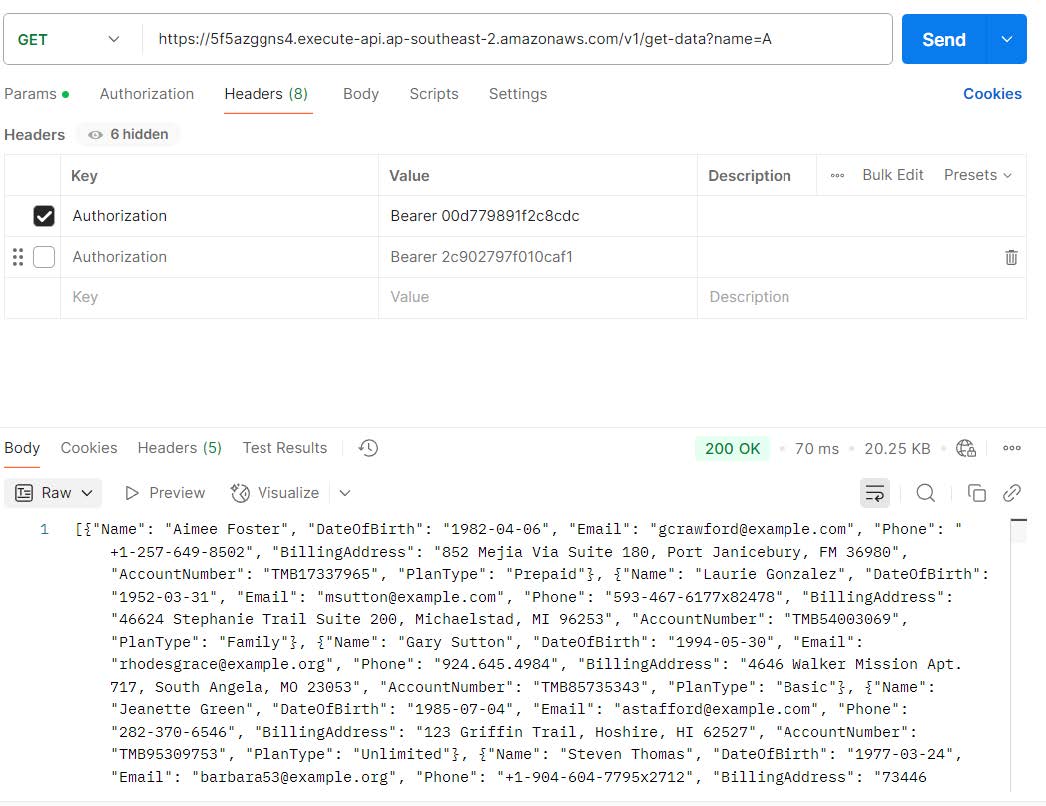}
	\caption{Admin Permission Test Result}
	\label{fig:admin-test-result}
	\end{minipage}
\end{figure}

All API requests must be accompanied by a valid Access Token. After the user logs in with the correct username and password, the generate\_unique\_hashed\_token function generates a unique 16-hexadecimal hash token, as shown in the code \ref{lst:encrypt}. This token is verified uniformly through the API Gateway and combined with RBAC policies to ensure that users can only access resources within their authorised scope. For example, regular users can only view their information, while administrators can view information about anyone. We have designed an example of this in AWS. The test is shown in Figures \ref{fig:user-test-result} and \ref{fig:admin-test-result}. 

\begin{lstlisting}[caption={Generate Unique Hashed Token Code}, label={lst:encrypt}]
	def generate_unique_hashed_token():
	while True:
	random_bytes = secrets.token_bytes(32)
	token = hashlib.sha256(random_bytes).hexdigest()[:16]
	if token not in SESSION_TOKENS:
	SESSION_TOKENS.add(token)
	return token
\end{lstlisting}

\noindent\textit{B. Optimise the Token Life Cycle}

Set the access token validity period to less than 1 hour and immediately destroy the current valid Token of the user when the user logs out or the permission is changed. 

\noindent\textit{C. Reasonable Rate-Limiting Strategy}

Set API call frequency limit for all user roles, no more than 100 API calls per minute. If the number of API calls exceeds 100, all the corresponding users' permissions will be blocked for 1 day, and an exception alarm will be triggered. Strengthen the log recording and audit mechanism: shorten the log audit cycle to once every half a month, and at least one "investigate any anomaly" principle should be adopted for each six audits to reduce the situation of missing abnormal logs. 

\noindent\textit{D. Implement HTTPS Encryption Transmission Policy}

Check and ensure that all APIs are equipped with TLS 1.2 or higher to ensure the confidentiality and integrity of data during transmission. 

\noindent\textit{E. Configuration Version Number}

Add the new version number v2 to the routing configuration, update the routing path, and modify the API document to maintain the compatibility of the old version and ensure the smooth transition of functions during the system upgrade. 

\subsection{Unified Access Security}
\label{Unified Access Security}

\subsubsection{Zero Trust Architecture (ZTA)}
\label{subsubsec:Zero Trust Architecture (ZTA)}
The margin-established protection framework assumes that all users and devices within the web can be trusted by default. However, this premise is no longer feasible. Based on the situation, the first key recommendation for improving access security is the adoption of a ZTA in the whole company. Moreover, various investigations and disclosed information all indicate that T-Mobile did not implement the ZTA before the 2021 data breach incident\cite{cso2020tmobilefcc}\cite{pcmag2020tmobilefcc}. Introducing ZTA would have reduced the onslaught openness of the T-Mobile environment significantly by guaranteeing rigorous confirmation and restricting lateral motion once inside the web.  In a zero-reliance framework, every exploiter, gimmick, and constituent of the web must be endlessly verified before being granted, irrespective of whether it is indoors or outside the organisational web. 

\subsubsection{Multi-Factor Authentication (MFA)}
\label{subsubsec:Multi-Factor Authentication (MFA)}
MFA is another layer of incorporated protection. Before 2021, the deficiency of compulsory MFA for all sensitive histories exposed companies like T-Mobile to inevitable hazards\cite{cso2020tmobilefcc}\cite{pcmag2020tmobilefcc}. MFA is an extra layer of protection beyond the watchword, making it significantly more difficult for wildcat users to derive even if the watchword has been compromised. Prioritise the consummate deployment of MFA for procuring inside and reducing the trust in individual-component hallmark. 

\subsubsection{Access Audit \& Cleanup}
\label{subsubsec:Access Audit & Cleanup}
Nonoperational, unneeded, or falsely configured histories accumulate over time and are possible introduction points for attackers. Before 2021, many companies, including T-Mobile, lacked the resources to place and rectify such cold histories\cite{aembit2023owasp}\cite{securityboulevard2021tmobile}\cite{strongdm2021tmobile}. A veritable audit and taxonomic clean-up of historical information is indispensable for potent administration. By conducting frequent reappraisals and quickly deactivating fresh certificates, we guarantee that only authorised people have access to the web and specific subnets in the company, thereby significantly strengthening the overall protection position. Additionally, use a reliable password generator for the device's service passwords, such as SSH, and then store the passwords properly. 

\subsection{Infrastructure Resilience \& Threat Monitoring}
\label{subsec:Infrastructure Resilience & Threat Monitoring}

\subsubsection{Endpoint Detection \& Response (EDR)}
\label{subsubsec:Endpoint Detection & Response (EDR)}
In 2021, many incidents escalated due to the deficiency of effective termination monitoring. EDR tools are also indispensable. EDR tools supply the existing clip profile into termination activities and enable speedy menace sensing, probing, and reaction. The EDR scheme helps to place and mitigate the menace before it can do widespread harm. 

\subsubsection{Proactive Threat Intelligence \& SIEM}
\label{subsubsec:Proactive Threat Intelligence & SIEM}
In combination with a Security Information and Event Management (SIEM)\footnote{Security Information and Event Management (SIEM) is a cybersecurity solution that aggregates and analyses log data from across an organization's IT infrastructure to detect, investigate, and respond to security threats in real time. By correlating events and applying analytics, SIEM enhances visibility, supports compliance reporting, and streamlines incident response through centralized dashboards and automated alerts\cite{ibm_siem}.} scheme, organisations can accomplish uninterrupted log aggregation, correlation, and anomaly sensing across their entire infrastructure. With a mature SIEM answer, they can not only heighten their profile, but also back former incident sensing and reaction, which was a major failing in the old breaches of T-Mobile. The subscription to the proactive menace news provider lets organisations prepare for known onslaught techniques and emerging threats. Additionally, the company should launch a bug bounty program to attract ethical hackers to find vulnerabilities, promptly fix verified issues, and provide appropriate rewards to vulnerability reporters\cite{ncioreview2024tmobile}\cite{itsecuritydemand2024cti}. 

\subsection{Data Protection \& Recovery}
\label{subsec:Data Protection & Recovery}

\subsubsection{Encryption \& Data Governance}
\label{subsubsec:Encryption & Data Governance}
Information categorisation systems must be implemented to ensure that information is managed decently according to its sensitivity and that controls are used appropriately. Sensible information must be coded at the remainder and in theodolite using potent, manufacturer-received cryptanalytic algorithms. Hapless encoding practices have historically led to terrible information escape in many organisations. Further hazards can be minimised by veritable information cleansing and information tagging procedures. This reduces the amount of excess or superannuated information and restricts the possible wallop of information wetting. 

\subsubsection{Incident Response \& Disaster Recovery}
\label{subsubsec:Incident Response & Disaster Recovery}
To trim the impact of breaches, it is critical to find a comprehensive and regularly tested incident reaction program. These plans must specify open functions, escalation methods, and communication schemes. In improver, standing, and catastrophe convalescence protocols should be strictly tested under naturalistic weather to guarantee that operations can rapidly recover from information deprivation or scheme compromises. 

\subsection{Security Culture \& Training}
\label{subsec:Security Culture & Training}

\subsubsection{Ongoing Cybersecurity Training}
\label{subsubsec:Ongoing Cybersecurity Training}
Veritable phishing and societal technology simulations are effective in raising consciousness and reducing the likelihood of human mistakes leading to compromise. Continual protection preparation is necessary for all employees, particularly those with eminent-prerogative functions. 

\subsubsection{Security Accountability}
\label{subsubsec:Security Accountability}
Eventually, protection of possession must be distributed across all departments. A potent civilisation of shared duty guarantees that protection is perceived not just as an IT care, but as a cardinal concern. The protection execution prosodies should be integrated into the faculty rating, which makes accountability and incentives for pro-fighting protection conduct. 

\subsection{Alternative Solutions and Options}

\begin{enumerate}[left=1em]
	\item \textbf{Use of Hosted API Security Services:} To reduce labour and technical overhead, implement API security gateways provided by cloud vendors such as AWS API Gateway to ensure authentication, authorisation, and traffic control.
	
	\item \textbf{Deployment of Open-Source Security Tools:} Use open-source security tools, such as deploying a WAF (Web Application Firewall)\footnote{A Web Application Firewall (WAF) is a security solution that monitors, filters, and blocks HTTP/HTTPS traffic between a web application and the internet to protect against common threats such as SQL injection, cross-site scripting (XSS), and file inclusion. Operating at the application layer (OSI Layer 7), WAFs can be deployed as network-based, host-based, or cloud-based solutions and are essential for safeguarding APIs, websites, and mobile apps from both known and emerging vulnerabilities\cite{wikipedia_waf}.} and combining it with the open-source log analysis platform ELK Stack\footnote{The ELK Stack—comprising Elasticsearch, Logstash, and Kibana—is an open-source suite widely adopted for centralised logging, real-time analytics, and visualising operational data. Elasticsearch handles search and indexing, Logstash ingests and transforms data from multiple sources, and Kibana provides interactive dashboards for monitoring and analysis.} for simple anomaly monitoring to reduce the cost of log input.
	
	\item \textbf{Phased Implementation of Security Measures:} Apply strict authentication and rate-limiting policies selectively on known vulnerable API endpoints as a staged approach to minimise financial impact while improving security posture.
\end{enumerate}

\subsection{Financial Analysis and Budget Breakdown}
\label{subsec:Financial Analysis and Budget Breakdown}
According to the above-mentioned security improvement plan, we have estimated the budgets for the improvement of each plan as shown in the table \ref{tab:budget-summary}: 

\begin{table}[htbp]
	\centering
	\caption{Estimated Budgets for IT Projects}
	\label{tab:budget-summary}
	\renewcommand{\arraystretch}{1.3}
	\begin{tabular}{|>{\centering\arraybackslash}m{4cm}|>{\centering\arraybackslash}m{2.5cm}|>{\centering\arraybackslash}m{2.5cm}|>{\arraybackslash}m{6.5cm}|}
		\hline
		\textbf{IT Project} & \textbf{Initial Budget} & \textbf{Budget (year)} & \textbf{Description} \\
		\hline
		API Gateway Deployment and Maintenance & \$7,500 & \$7,500 & Purchase or subscribe to a hosted API management service (such as Azure API Management, AWS API Gateway) for unified Access Token authentication, routing management, and Rate Limiting control. \\
		\hline
		Token Generation and Authentication Mechanism Development & \$3,000 & \$3,000 & Design and implement access token and refresh token mechanisms, including the development and deployment of the \texttt{generate\_unique\_hashed\_token} function. \\
		\hline
		Firewall Deployment (Palo Alto PA-7500) & \$900,000 & \$10,000 & Deploy multiple PA-7500 firewall devices when considering load balancing at the enterprise boundary network. \\
		\hline
		Devices Update & \$500,000 & \$20,000 & Update the old equipment in the company that needs to be phased out (devices with security vulnerabilities or not supporting the latest security protocols). \\
		\hline
		Hiring Pros for Network and Computer Device Settings & \$1,000,000 & \$600,000 & Hiring staff holding professional certificates, such as CCNP-Security certificate, to perform security maintenance and configuration for the entire network. \\
		\hline
		Bug Bounty Program & \$100,000 & \$100,000 & Post bounty programs on ethical hacking platforms offering appropriate bonuses for valid vulnerabilities. \\
		\hline
		Rate Limiting System Configuration & \$1,500 & \$1,500 & Configure the call frequency restriction policy based on user roles and integrate the alarm trigger and account ban logic. \\
		\hline
		Log Audit Systems and SIEM Tools & \$4,000 & \$4,000 & Purchase and deploy a security information and event management system to centrally collect, store, and analyse API access logs and support anomaly detection. \\
		\hline
		HTTPS Encryption Upgrade and Certificate Management & \$800 & \$800 & Configure TLS 1.2 or higher version encryption and update the SSL certificate regularly to ensure communication security. \\
		\hline
		Version Control and API Documentation Updates & \$1,000 & \$1,000 & Design a new route version v2, update and maintain API documentation, and ensure consistency and compatibility during development and use. \\
		\hline
	\end{tabular}
\end{table}

\begin{figure}[htbp]
	\centering
	\includegraphics[width=0.8\linewidth]{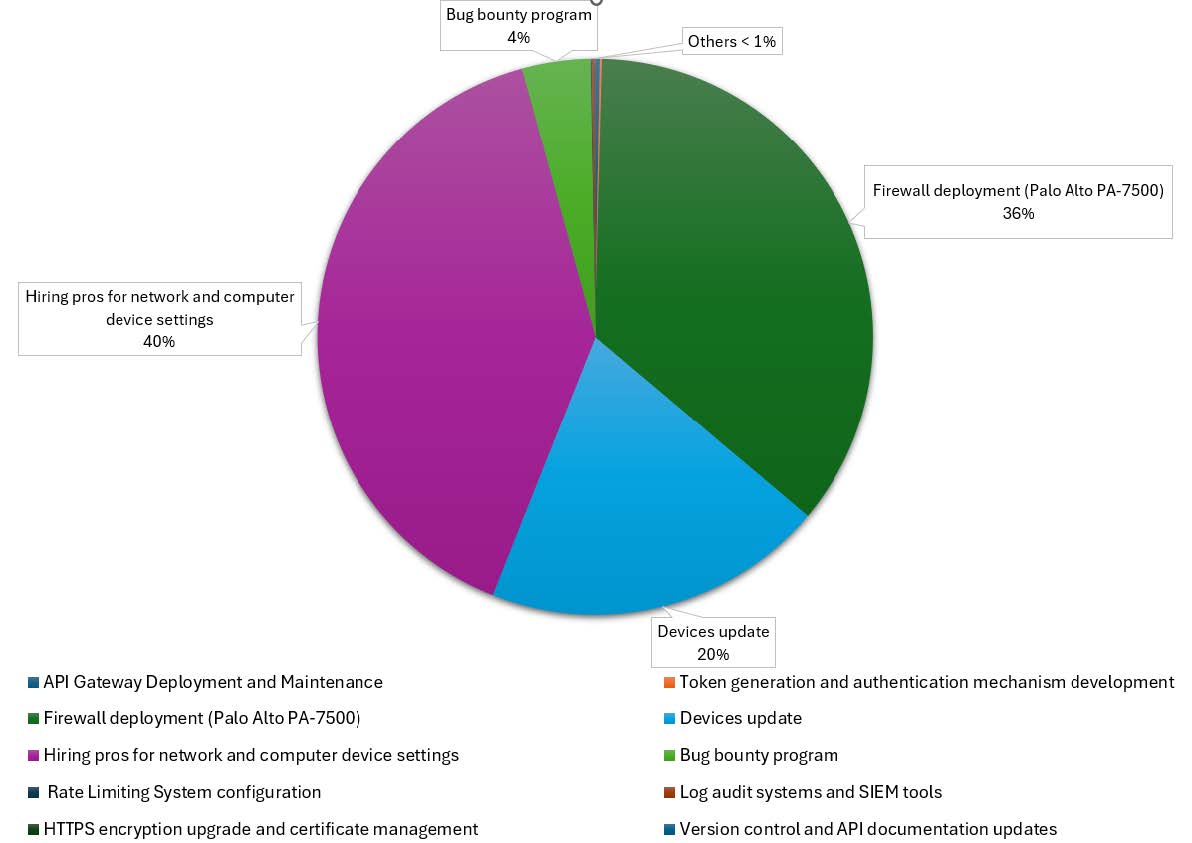}
	\caption{The Initial Budget Proportion of Each IT Project}
	\label{fig:budget-proportion}
\end{figure}

Initial investment: approximately \$2,517,800. The budget proportion chart of each IT project is shown in Figure \ref{fig:budget-proportion}.

\subsection{ROI Justification}
\label{ROI Justification}
An initial investment of \$2,517,800 and approximately \$747,800 annually in API-related equipment
maintenance and upgrades can significantly reduce the risk of large-scale data breaches. According
to IBM's 2023 Cost of Data Breaches Report and the latest results in 2021, the average direct
economic loss to a company per data breach is approximately \$535 million, including fines, legal
fees, customer loss, and brand reputation damage\cite{ibm2023breachreport}. In the data breaches of T-Mobile in 2021 and
2023, the direct economic loss is more than 350 million US dollars, and the indirect economic loss is
more than 150 million US dollars\cite{mobileidworld2021settlement}\cite{wired2023breachagain}. The total cost of the security investment over five years
is about \$5.509 million, which is only 1.03\% of the typical data breach loss.

\section{Comprehensive Security Audit and Vulnerability Assessment of T-Mobile Systems}
\label{sec:Comprehensive Security Audit and Vulnerability Assessment of T-Mobile Systems}

\subsection{Security Audit Overview}
\label{subsec:Security Audit Overview}

\subsubsection{Understanding Security Audits}
\label{subsubsec:Understanding Security Audits}

Security audit is a combination of manual inspection and automated scanning to systematically assess
an organisation's information systems, networks, and infrastructure to identify current vulnerabilities
and whether current security measures are effective against internal and external threats. The
existence of a security audit is essential in a modern IT environment for the following reasons:

\begin{enumerate}[left=1em]
	\item \textbf{Risk Identification:} The audit can identify the possible vulnerabilities of misconfiguration, improper firewall rules, software not updated in time, insufficient password strength, and too high user privileges, and the organisation can take timely repair measures to reduce the attack surface effectively.
	
	\item \textbf{Regulatory Compliance:} ISO/IEC 27001, PCI-DSS, and GDPR emphasise that regular security audit is an important way to prove the compliance of organisations.
	
	\item \textbf{Prevent Security Incidents:} Actively detecting vulnerabilities in the system and fixing them can reduce the risk of security incidents such as data leakage and system interruption.
	
	\item \textbf{Enhance Stakeholder Confidence:} The act of conducting regular security audits and fixing vulnerabilities is more likely to earn the trust of customers, investors, and regulators.
\end{enumerate}

\subsubsection{Scope of the Audit}
\label{subsubsec:Scope of the Audit}
Based on the security solution strategies and plans discussed in the previous sections, we will now test whether T-Mobile has implemented these security strategies. The scope of this security audit includes public-facing web infrastructure, backend services, and 5G hardware products. Due to geographical reasons and equipment limitations, we can't test the wireless signals, surrounding facilities, and actual hardware devices near the T-Mobile headquarters (physical devices still need to be obtained for testing). However, because of geographical factors and the narrow attack surface, these are often the most easily overlooked areas with security vulnerabilities. 

The tools and platforms used in this security audit are Nmap\footnote{A powerful open-source network scanner used for host discovery, port scanning, and service enumeration. It supports OS fingerprinting and scripting for vulnerability detection.}, Burp Suite\footnote{A comprehensive web application security testing platform that includes tools for intercepting traffic, scanning for vulnerabilities, and automating attacks like XSS and SQL injection.}, sqlmap\footnote{An automated penetration testing tool that detects and exploits SQL injection vulnerabilities. It supports database fingerprinting, data extraction, and remote command execution.}, hydra\footnote{A fast and flexible login cracker that supports numerous protocols (e.g., SSH, FTP, HTTP) for brute-force password attacks against remote authentication services.}, gobuster\footnote{A directory and file brute-forcing tool written in Go. It is commonly used to discover hidden web paths, subdomains, and virtual hosts using wordlists.}, dirsearch\footnote{A command-line tool designed to brute-force directories and files on web servers. It supports recursive scanning, custom headers, and extension filtering.}, waybackurls\footnote{A utility that fetches historical URLs for a domain from the Wayback Machine (Internet Archive), aiding in web reconnaissance and endpoint discovery.}, binwalk\footnote{A firmware analysis tool used to extract embedded files and reverse-engineer binary images. It is widely used in embedded systems and IoT security research.}, and Shodan\footnote{A search engine for Internet-connected devices. It indexes metadata from services like HTTP, SSH, and Telnet, enabling researchers to discover exposed systems and vulnerabilities.}. 

\subsection{Identification of System Vulnerabilities}
\label{subsec:Identification of System Vulnerabilities}

\subsubsection{Web Infrastructure Assessment}
\label{subsubsec:Web Infrastructure Assessment}

\begin{figure}[htbp]
	\centering
	\includegraphics[width=0.8\linewidth]{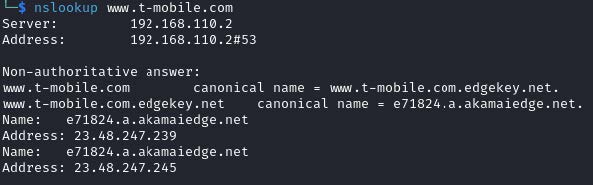}
	\caption{Using nslookup to Query the IP Address of T-Mobile}
	\label{fig:query-IP}
\end{figure}

\begin{figure}[htbp]
	\centering
	\includegraphics[width=0.8\linewidth]{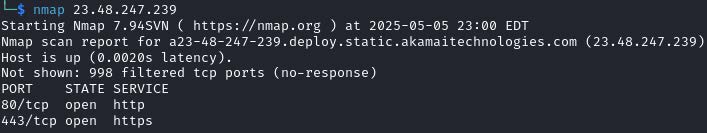}
	\caption{Nmap Scan Result1}
	\label{fig:nmap-scan-result1}
\end{figure}

\begin{figure}[htbp]
	\centering
	\includegraphics[width=0.8\linewidth]{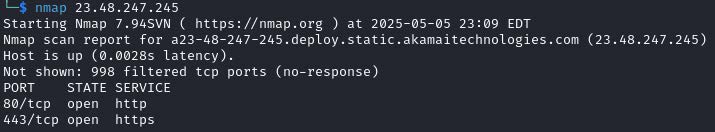}
	\caption{Nmap Scan Result2}
	\label{fig:nmap-scan-result2}
\end{figure}

The date of the security audit is in early May 2025. Overall, no significant problems were found in the security audit of the Web Infrastructure. The test system we use in this audit is Kali, a well-known Debian security system with numerous security auditing tools. First, we use nslookup\footnote{nslookup is a command-line network utility used to query Domain Name System (DNS) servers for information about domain names and IP addresses. It supports both forward and reverse lookups, allowing users to retrieve DNS records such as A, MX, NS, and PTR. Commonly used for troubleshooting DNS issues, nslookup operates in both interactive and non-interactive modes across Windows, Linux, and macOS platforms.} to query the IP address of T-Mobile's official website. The query situation is shown in Figure \ref{fig:query-IP}. Then, we use Nmap to scan the IP address to check its port and service availability, as shown in Figure \ref{fig:nmap-scan-result1} and Figure \ref{fig:nmap-scan-result2}. 

\begin{figure}[htbp]
	\centering
	\includegraphics[width=0.8\linewidth]{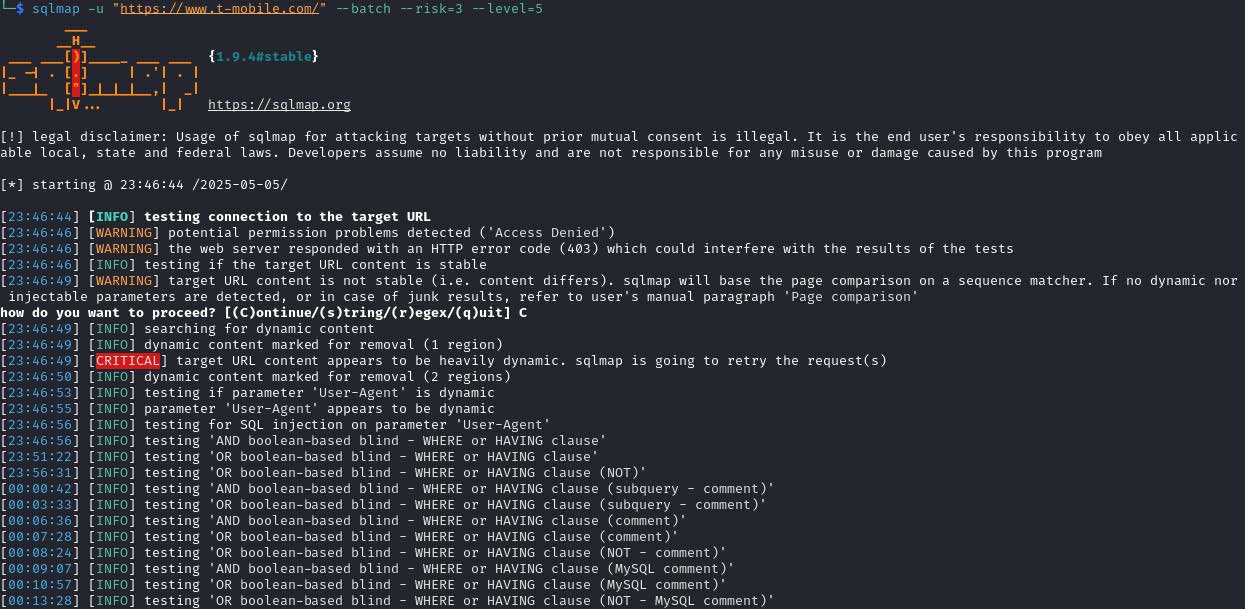}
	\caption{sqlmap Scan Result}
	\label{fig:sqlmap-scan-results}
\end{figure}

According to the scanning results, only HTTP and HTTPS services are available for both IP addresses, and there are no management services such as SSH (port 22) or VNC\footnote{Virtual Network Computing (VNC) is a remote desktop sharing protocol that enables users to view and control another computer's graphical interface over a network. It operates using the Remote Frame Buffer (RFB) protocol and is widely used for remote support, system administration, and cross-platform desktop access.} (port 5900). From this, it can be seen that no obvious commonly used management or other service ports are exposed on public servers. Connecting to a VPN or the company's internal network may be necessary to provide other services, which is excellent work from a security perspective. After that, we used sqlmap for blind scanning. Again, no suitable injection points or vulnerabilities like XSS were found. The scanning results are shown in Figure \ref{fig:sqlmap-scan-results}. 

\begin{figure}[htbp]
	\centering
	\includegraphics[width=0.8\linewidth]{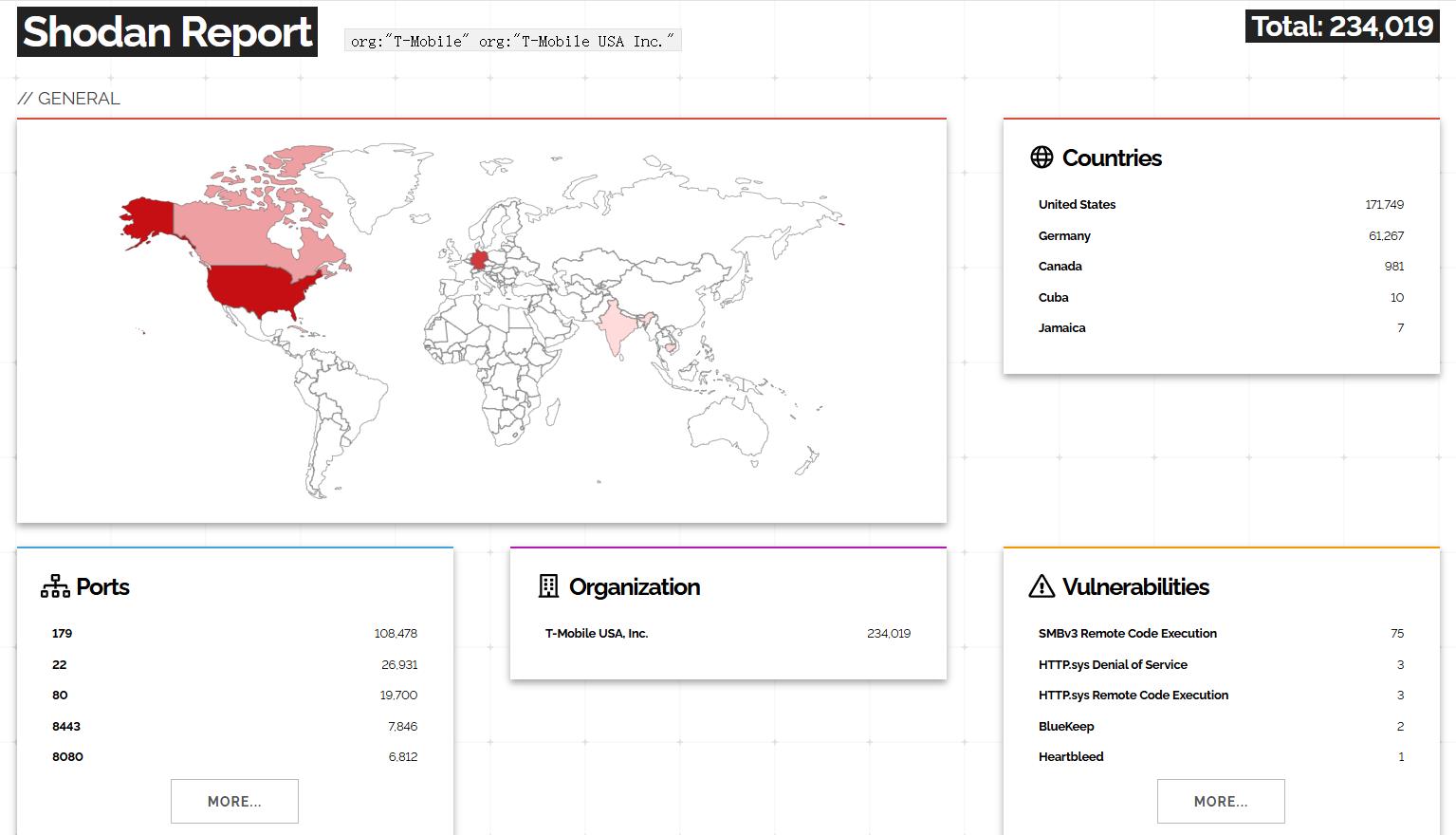}
	\caption{Shodan Search Report on T-Mobile Devices}
	\label{fig:shodan-search-report}
\end{figure}

Next, we use Shodan to query the Internet for T-Mobile devices that may have vulnerabilities and display their IP addresses, as shown in Figure \ref{fig:shodan-search-report}. The search report indicates 84 possible vulnerabilities, including SMBv3 Remote Code Execution(CVE-2020-0796\footnote{CVE-2020-0796, also known as SMBGhost, is a critical remote code execution vulnerability in Microsoft’s Server Message Block version 3.1.1 (SMBv3) protocol. Disclosed in March 2020, it affects Windows 10 and Windows Server versions 1903 and 1909. The flaw stems from improper handling of compressed data packets, allowing unauthenticated attackers to execute arbitrary code on vulnerable systems. Classified as "wormable", it poses a high risk of self-propagating malware like EternalBlue, prompting urgent patching and mitigation efforts by Microsoft and global cybersecurity agencies\cite{cve20200796}.}): 75, HTTP.sys Denial of Service: 3, HTTP.sys Remote Code Execution(CVE-2015-1635\footnote{CVE-2015-1635 is a critical remote code execution vulnerability in Microsoft’s HTTP.sys driver, disclosed in April 2015 and addressed by security bulletin MS15-034. It affects Windows 7 SP1, Windows Server 2008 R2 SP1, Windows 8/8.1, and Windows Server 2012, allowing attackers to send specially crafted HTTP requests that trigger buffer overflows and potentially execute arbitrary code with SYSTEM privileges. Widely exploited and assigned a CVSS score of 10.0, this flaw underscores the importance of timely patching and robust vulnerability management\cite{cve20151635}.}): 3, BlueKeep\footnote{The BlueKeep vulnerability (CVE-2019-0708) is a critical remote code execution flaw in Microsoft’s Remote Desktop Protocol (RDP), affecting older Windows versions such as XP, Vista, 7, and Server 2003/2008. Discovered in May 2019, it is classified as “wormable,” meaning it can self-propagate across vulnerable systems without user interaction, like the WannaCry ransomware. Microsoft and global cybersecurity agencies, including the NSA and CISA, urged immediate patching due to its potential for widespread exploitation\cite{wikipediaBlueKeep}.}: 2, and Heartbleed\footnote{The Heartbleed vulnerability (CVE-2014-0160) is a critical flaw in specific versions of the OpenSSL cryptographic library, publicly disclosed in April 2014. It stems from improper bounds checking in the TLS heartbeat extension, allowing attackers to read up to 64KB of server memory per request. This exposed sensitive data such as private keys, passwords, and session tokens, enabling eavesdropping and impersonation. Heartbleed affected millions of systems globally and prompted widespread patching, certificate revocation, and scrutiny of open-source security practices\cite{wikipediaHeartbleed}.}: 1. Assuming these devices all belong to T-Mobile. 

\begin{figure}[htbp]
	\centering
	\includegraphics[width=0.8\linewidth]{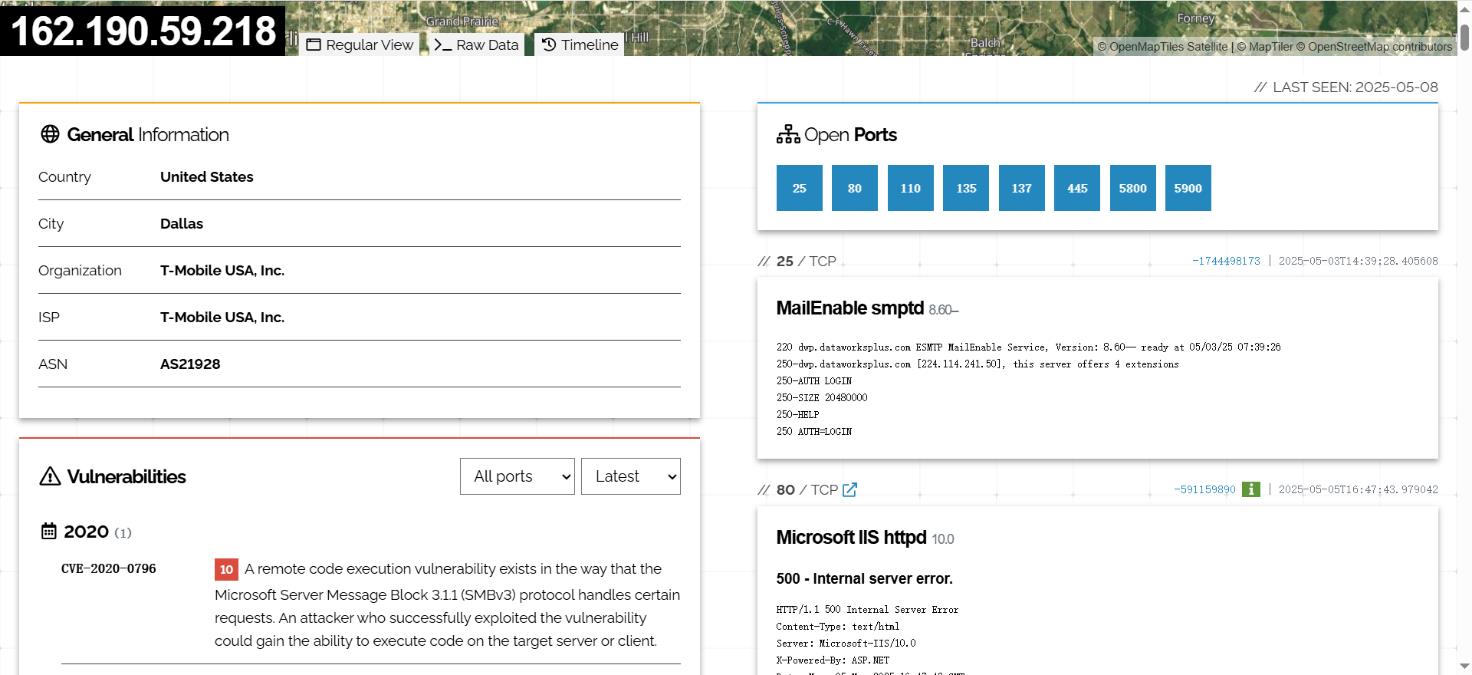}
	\caption{Device Detail in Shodan}
	\label{fig:device-detail}
\end{figure}

\begin{figure}[htbp]
	\centering
	\includegraphics[width=0.8\linewidth]{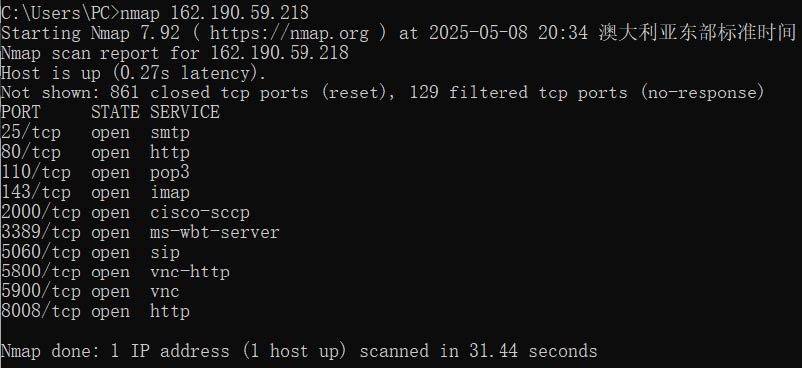}
	\caption{Nmap Scan Result3}
	\label{fig:nmap-scan-result3}
\end{figure}

\begin{figure}[htbp]
	\centering
	\includegraphics[width=0.8\linewidth]{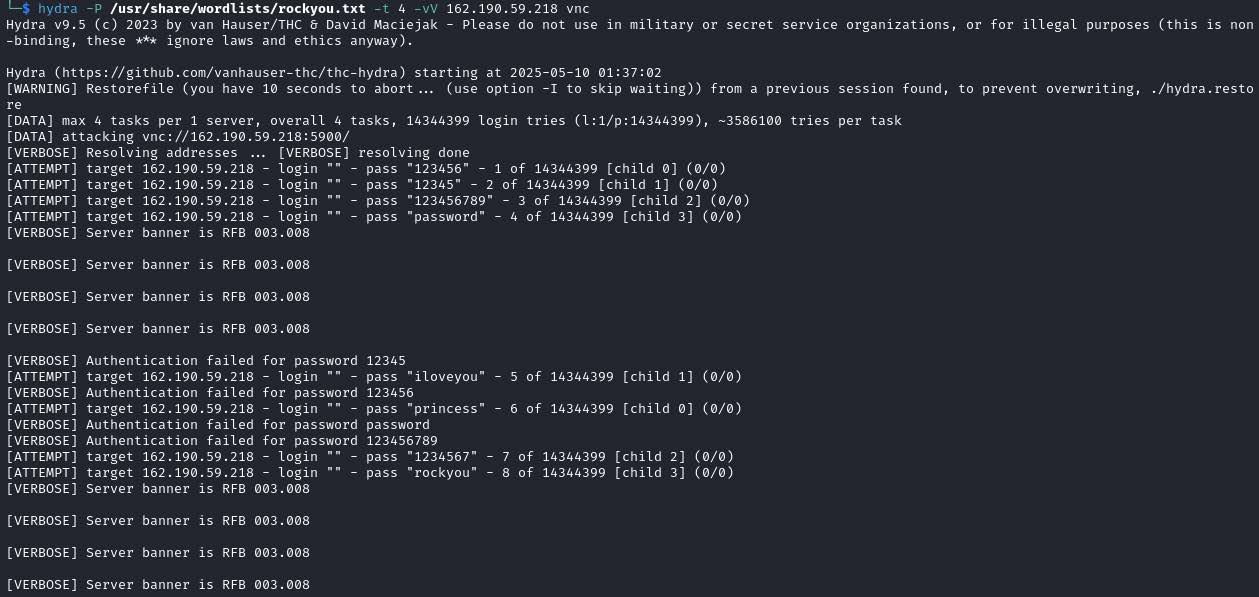}
	\caption{hydra Brute Force}
	\label{fig:hydra-brute-force}
\end{figure}

We have chosen a device with SMBv3 Remote Code Execution in Figure \ref{fig:device-detail}, and we scan and check the device's details. After Nmap scanning in Figure \ref{fig:nmap-scan-result3}, it seems like this device has already closed the SMB service port, but we can also recognise that it has opened the VNC service, so we use VNC Viewer to try to connect to it. Then we found out it has a connection password, so we decided to use hydra and rockyou.txt as a dictionary to brute-force the password shown in Figure \ref{fig:hydra-brute-force}. 

\begin{figure}[htbp]
	\centering
	\includegraphics[width=0.8\linewidth]{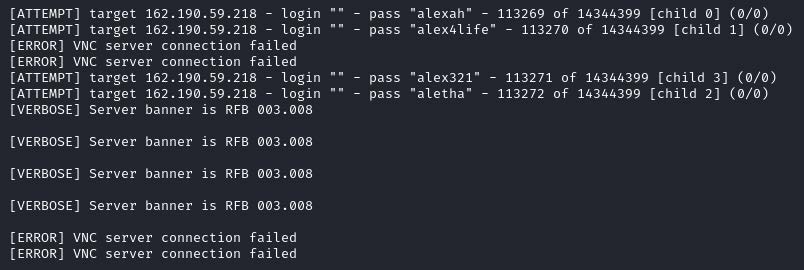}
	\caption{hydra Brute Force Result}
	\label{fig:hydra-brute-force-result}
\end{figure}

The hydra result is shown in Figure \ref{fig:hydra-brute-force-result}. It can be seen that we failed to crack the password using rockyou.txt, due to the limitations of each cracked thread and VNC password attempt, the cracking time was too long. Thus, we didn't attempt the entire rockyou.txt dictionary, instead focusing on more than 110,000 passwords. This VNC didn't use TLS or encryption and implemented limitations to protect it from brute force attacks. 

We then selected devices with T-Mobile identification from Shodan's CVE\footnote{Common Vulnerabilities and Exposures (CVE) is a publicly accessible database that provides standardised identifiers for known cybersecurity vulnerabilities in software and hardware systems. Managed by The MITRE Corporation and sponsored by the U.S. Cybersecurity and Infrastructure Security Agency (CISA), CVE enables consistent tracking, sharing, and remediation of security issues across tools, platforms, and organisations worldwide\cite{cveofficialsite}.} list for scanning, and the results are shown in Table \ref{tab:subnet-vlan}. Due to the geographical location issue of our access IP, some devices cannot be accessed or perform accurate scans. Moreover, Shodan is not a real-time scanning engine, and the results may not be correct. The precise results still need to be confirmed by manual scanning. This might be why the devices we selected and scanned found nothing, and no corresponding vulnerabilities were discovered. 

\begin{table}[htbp]
	\centering
	\caption{Subnet Division Based on VLAN}
	\label{tab:subnet-vlan}
	\begin{tabular}{|c|c|c|}
		\hline
		\textbf{Device IP} & \textbf{Vulnerability Shown in Shodan} & \textbf{Scan Result} \\
		\hline
		162.191.33.31     & CVE-2020-0796 ({SMBGhost}) & Host Down \\
		162.191.211.159   & CVE-2020-0796 ({SMBGhost}) & Host Down \\
		162.185.120.21    & CVE-2020-0796 ({SMBGhost}) & Port Down \\
		162.172.12.225    & CVE-2019-0708 ({BlueKeep}) & Port Down \\
		72.250.77.248     & CVE-2019-0708 ({BlueKeep}) & Host Down \\
		162.191.29.124    & CVE-2014-0160 ({Heartbleed}) & Vulnerability Not Available \\
		162.185.218.73    & CVE-2015-1635 & Vulnerability Not Available \\
		162.191.47.206    & CVE-2015-1635 & Host Down \\
		162.185.197.15    & CVE-2015-1635 & Host Down \\
		\hline
	\end{tabular}
\end{table}

\subsubsection{URL Enumeration and Further Assessment}
\label{subsubsec:URL Enumeration and Further Assessment}

\begin{figure}[htbp]
	\centering
	\includegraphics[width=0.8\linewidth]{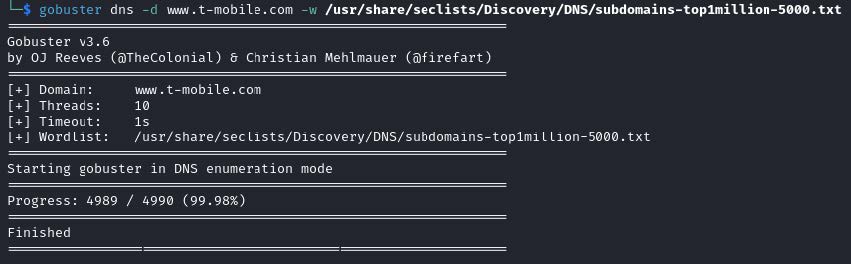}
	\caption{gobuster Scanning Result}
	\label{fig:gobuster-scanning-result}
\end{figure}

To conduct in-depth scanning and vulnerability finding, we use gobuster and dirsearch to look for sensitive directories (such as admin login page) and possible injection points, such as: "index.php?id=1". First, the scanning results using gobuster are shown in Figure \ref{fig:gobuster-scanning-result}. No sensitive directories or SQL injection points were found. 

\begin{figure}[htbp]
	\begin{minipage}[t]{0.48\textwidth}
		\centering
		\includegraphics[width=1\linewidth]{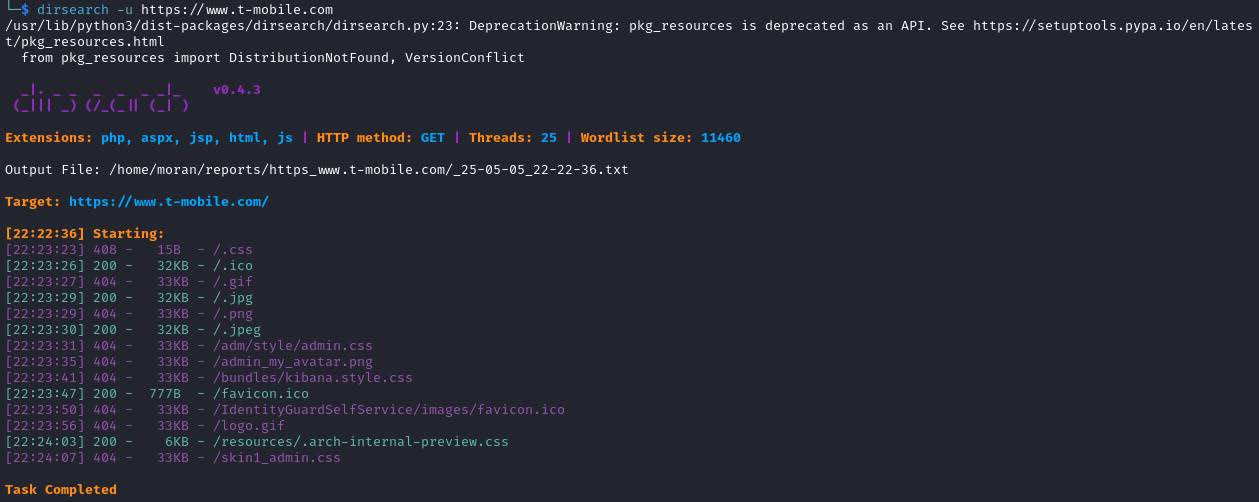}
		\caption{dirsearch Scanning Results}
		\label{fig:dirsearch-scanning-results}
	\end{minipage}
	\hfill
	\begin{minipage}[t]{0.48\textwidth}
	\centering
	\includegraphics[width=1\linewidth]{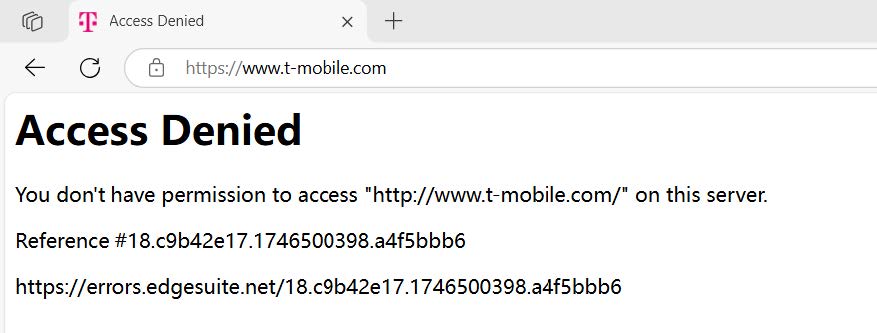}
	\caption{T-Mobile Temporarily Blocks Access}
	\label{fig:temporarily-blocks-access}
	\end{minipage}
\end{figure}

Then, we used a larger dictionary for brute-force scanning and dirsearch to conduct further URL scans. The results are shown in Figure \ref{fig:dirsearch-scanning-results}. No sensitive directories were exposed, and only some meaningless directories, perhaps standard endpoints, have been protected or obfuscated. It is worth noting that before and after the scan, we visited T-Mobile's official website and found measures like IPS for scanning on the official website. The requests of this IP have been temporarily prohibited, as shown in Figure \ref{fig:temporarily-blocks-access}. This measure can indeed effectively prevent such brute-force scans. 

\subsubsection{Archived Endpoint Analysis}
\label{subsubsec:Archived Endpoint Analysis}

\begin{figure}[htbp]
	\centering
	\includegraphics[width=0.45\linewidth]{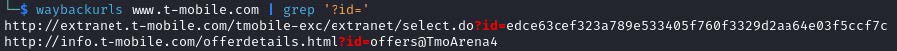}
	\includegraphics[width=0.45\linewidth]{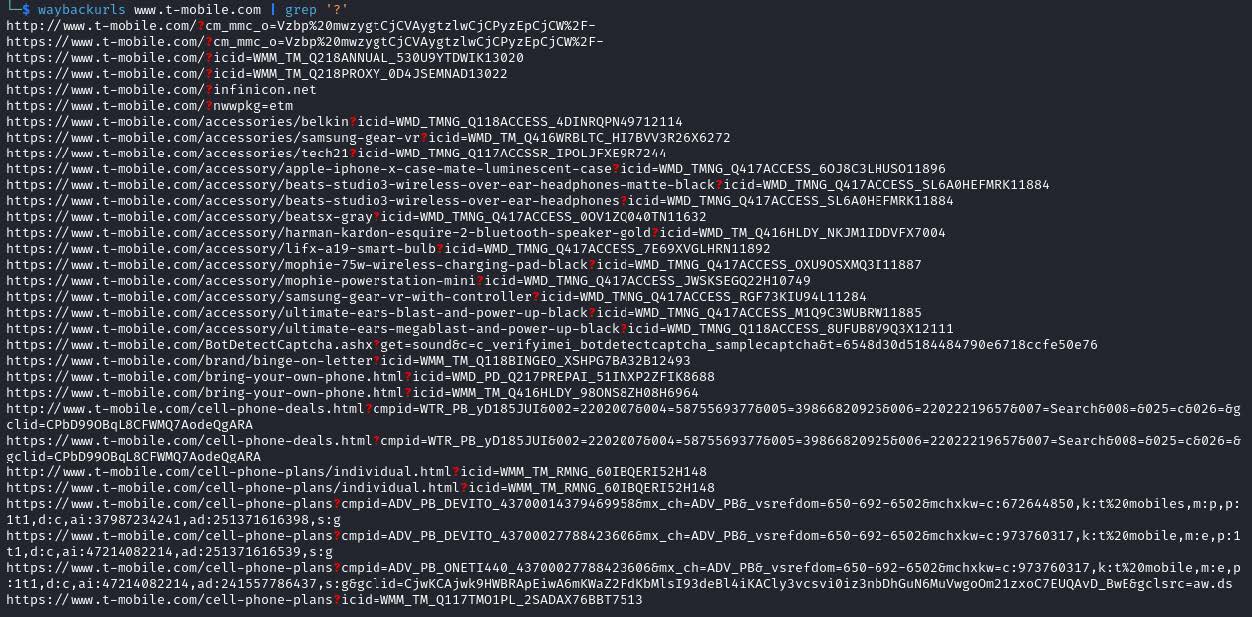}
	\caption{waybackurl Result}
	\label{fig:waybackurl-result}
\end{figure}

Since the directory brute force scanning failed to find meaningful injection points and sensitive directories, we decided to use wayback URLs to identify deprecated URLs and historical injection points. We set the filter URL to contain "?" and "?id=" because links containing this symbol will likely indicate vulnerabilities such as SQL injection or XSS. The results of using wayback URLs are shown in Figure \ref{fig:waybackurl-result}.

\begin{figure}[htbp]
	\centering
	\includegraphics[width=0.8\linewidth]{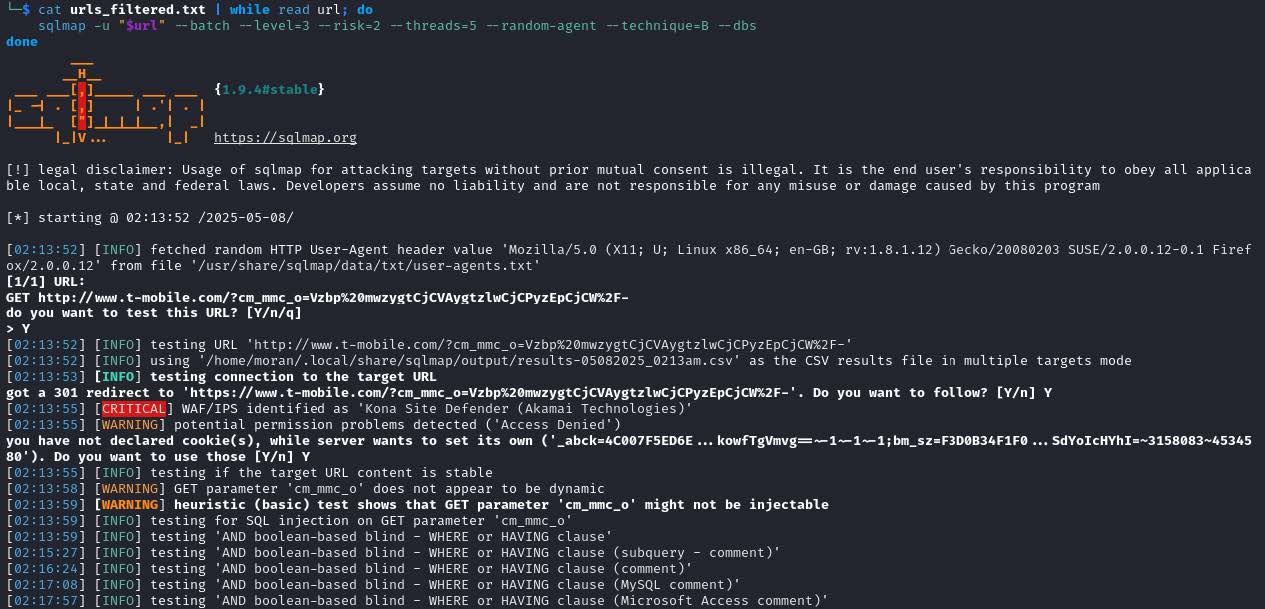}
	\caption{sqlmap Scanning Result}
	\label{fig:sqlmap-scanning-result}
\end{figure}

After seeing the result, we used the command: "echo "www.t-mobile.com" | waybackurls > urls.txt", to consolidate all results into urls.txt. Then use the command: "cat urls.txt | grep "?" | grep "=" | tee urls\_filtered.txt", to filter for the link contains "?". The link and save it to urls\_filtered.txt. Using sqlmap for scanning, part of the result is shown in Figure \ref{fig:sqlmap-scanning-result}. According to the final scanning results of sqlmap, the website does have a WAF or an IPS to prevent scanning. Facts have proven that T-Mobile strengthened its security after two data breach incidents. The WAF it uses is the Kona Site Defender\footnote{Kona Site Defender is Akamai’s cloud-based security solution that combines a Web Application Firewall (WAF) and DDoS protection to safeguard websites, APIs, and applications. It offers real-time threat mitigation, customizable rule sets, and advanced analytics, leveraging Akamai’s global edge network to detect and block attacks with minimal latency\cite{akamaiKona2023}.} product of Akamai Technologies. It combines automated DDoS mitigation with a highly scalable and accurate WAF and is designed to provide integrated website protection against DDoS and web application attacks. 

\begin{figure}[htbp]
	\centering
	\includegraphics[width=0.45\linewidth]{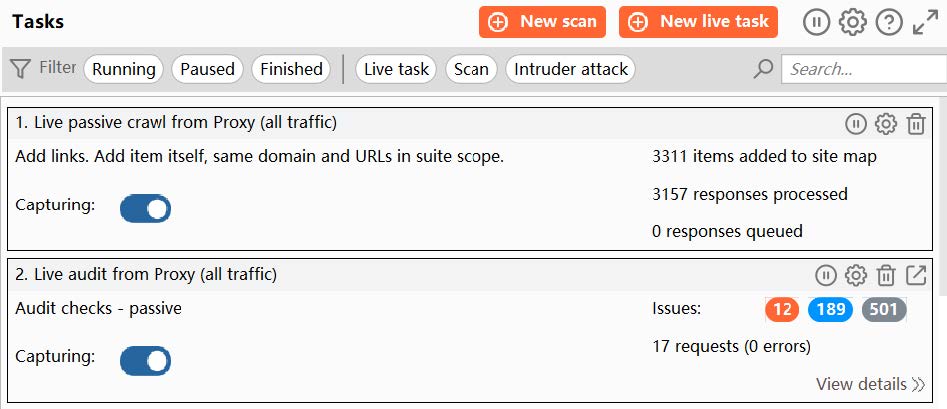}
	\includegraphics[width=0.45\linewidth]{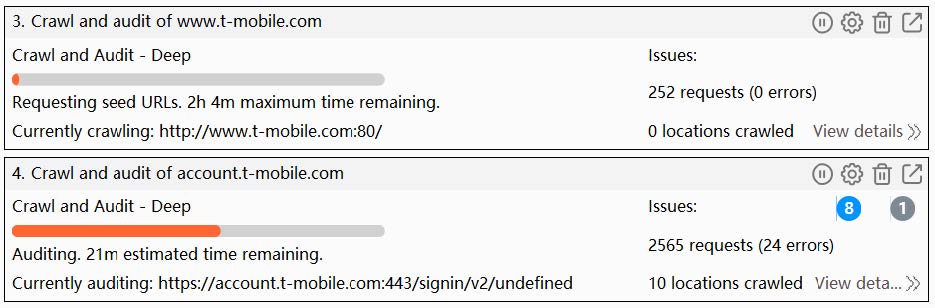}
	\caption{Scanning Targets}
	\label{fig:scanning-target}
\end{figure}

\begin{figure}[htbp]
	\begin{minipage}[t]{0.48\textwidth}
	\centering
	\includegraphics[width=1\linewidth]{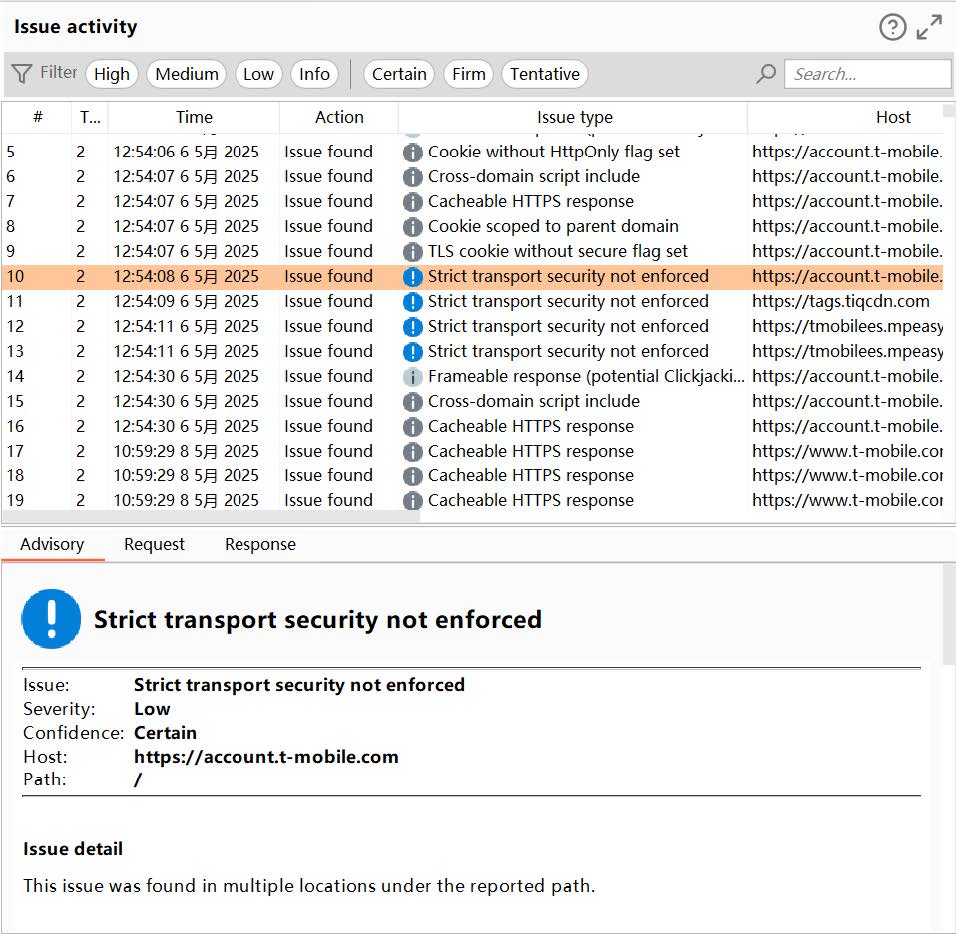}
	\caption{Scanning Results}
	\label{fig:scanning-results}
	\end{minipage}
	\hfill
	\begin{minipage}[t]{0.48\textwidth}
	\centering
	\includegraphics[width=1\linewidth]{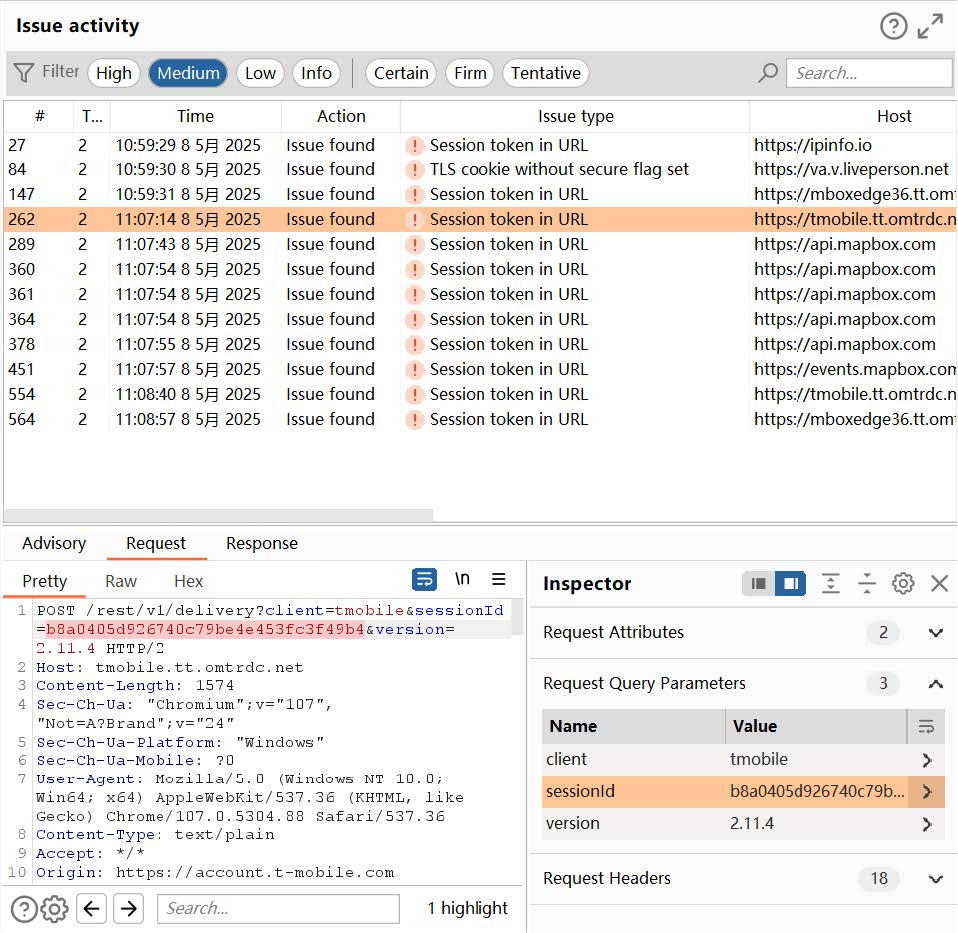}
	\caption{Medium Issues}
	\label{fig:medium-lssues}
	\end{minipage}
\end{figure}

After that, we attempted to use Burp Suite to conduct automatic and manual Web scans. The scanned URLs are shown in Figure \ref{fig:scanning-target}, and part of the results are shown in Figure \ref{fig:scanning-results}. According to the scanning results, it can be seen that the most common issue types are low and info types, with a small number of Medium Issues, as shown in Figure \ref{fig:medium-lssues}, and medium issues are mainly concentrated on the nested third-party web pages or resource websites of T-Mobile. No high-risk vulnerabilities were found in the results.

\begin{figure}[htbp]
	\begin{minipage}[t]{0.48\textwidth}
	\centering
	\includegraphics[width=1\textwidth]{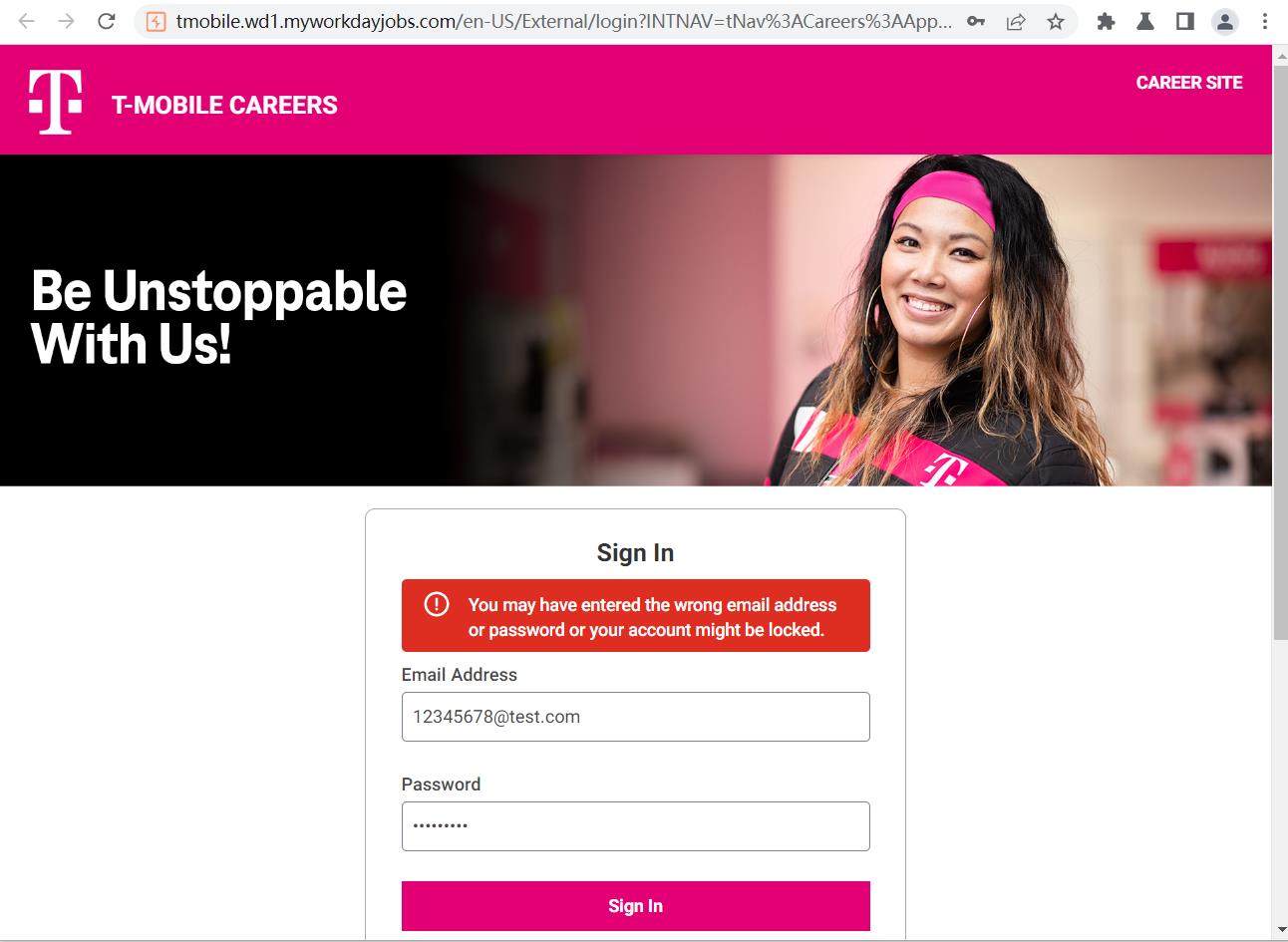}
	\caption{\href{https://tmobile.wd1.myworkdayjobs.com/en-US/External/login}{T-Mobile Careers Login Page}}
	\label{fig:login-page}
	\end{minipage}
	\hfill
	\begin{minipage}[t]{0.48\textwidth}
	\centering
	\includegraphics[width=0.8\linewidth]{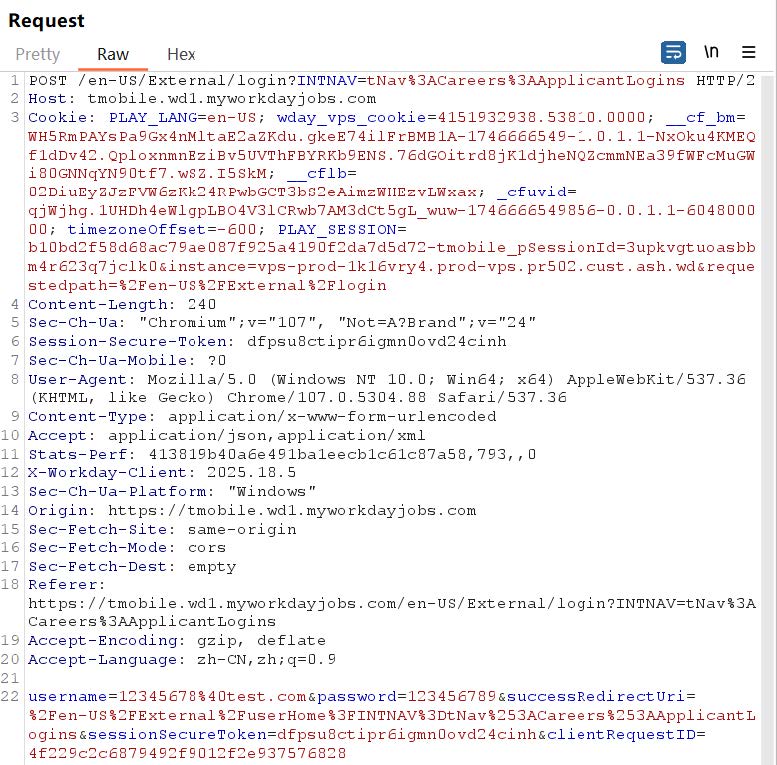}
	\caption{Captured Request in Burp Suite}
	\label{fig:captured-request}
	\end{minipage}
\end{figure}

Furthermore, we made an interesting discovery in the manual scan of Burp Suite. On the login page of 'T-Mobile Careers', which is shown in Figure \ref{fig:login-page}, we discovered a vulnerability for plaintext transmission of usernames and passwords. The requests and responses for the vulnerability are shown in Figure \ref{fig:captured-request}. We used the account number 12345678@test.com and the password 123456789 for the test. As can be seen in the figure, the requested information for the plaintext account password was captured (Using Burp Suite certificate and proxy, bypassing the TLS encryption). 

\subsubsection{5G Hardware and Firmware Security Research}
\label{subusbsec:5G Hardware and Firmware Security Research}

\begin{figure}[htbp]
	\centering
	\includegraphics[width=0.4\linewidth]{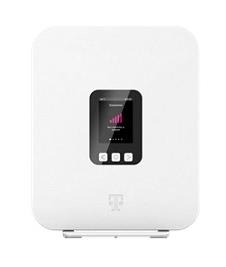}
	\includegraphics[width=0.4\linewidth]{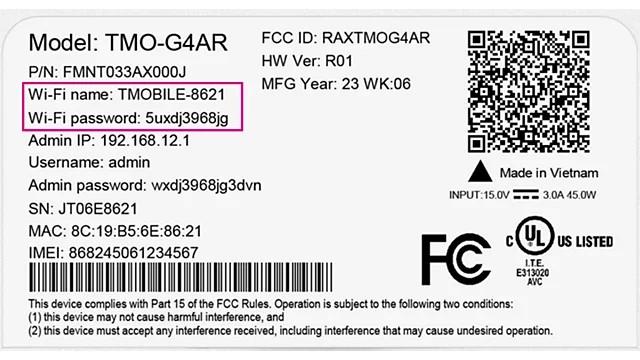}
	\includegraphics[width=0.4\linewidth]{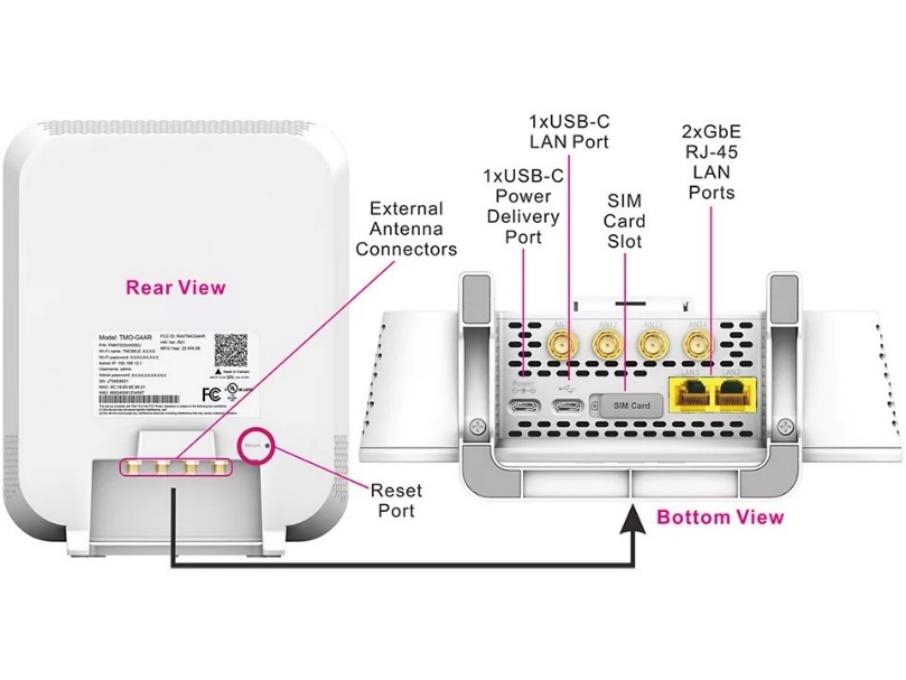}
	\caption{The Device}
	\label{fig:device}
\end{figure}

We chose a 5G home hardware device, as shown in Figures \ref{fig:device}. We've found that T-Mobile is currently very mindful of hardware security issues, like they may use pseudo-random 11-bit passwords to protect WiFi shown as Figure \ref{fig:device}, and 14-bit pseudo-random passwords to protect administrator accounts, this looks pretty good, but since we don't have the device, we can't be sure that the login page or firmware has no bugs. The password of 11 lowercase letters and numbers will have 131,621,703,842,267,136 combinations, and it will take about 318 days to crack this WiFi password (WPA/WPA2) using GPU-RTX4090 in advanced cracking software called hashcat\footnote{Hashcat is an open-source password recovery tool renowned for its speed, flexibility, and support for over 300 hashing algorithms. Designed for ethical hacking and security auditing, it leverages CPU, GPU, and other hardware accelerators to perform brute-force, dictionary, mask, and hybrid attacks. Hashcat is widely used in penetration testing and forensic analysis to assess password strength and uncover vulnerabilities in authentication systems\cite{hashcat}.}, for the individual computer, which is infeasible. In addition, the gateway uses an automatic update method, which cannot be manually updated, and there is no web page and offline update method to download firmware, which is also good from the point of view of device security, which prevents hackers from reverse-engineering and analysing the firmware of the device without the device. However, this method will not work if the device needs to be updated offline in case of a problem. According to the current data on the official website, the device is relatively safe, but the specific situation also requires the real device to confirm whether it is secure. We can conclude that even though T-Mobile's 5G gateway devices disabled their offline update firmware feature, it could still lead to supply chain attacks. The fact is that while T-Mobile does its best to protect their firmware files from being exposed to the Internet, the attacker can still using reverse engineering and modify the firmware by purchasing the device and dumping the firmware from the device's flash chip, and discover its firmware update methods and requested APIs in the firmware, and, In T-Mobile's history, due to their initial contact-proof protection of device firmware, likely, the update address and API where the device updates its firmware will may not be protected. The attacker could hack into its API and firmware file storage server and insert malicious code into the original firmware, then initiate an active update to all T-Mobile devices containing the malicious code. 

\begin{figure}[htbp]
	\centering
	\includegraphics[width=0.8\linewidth]{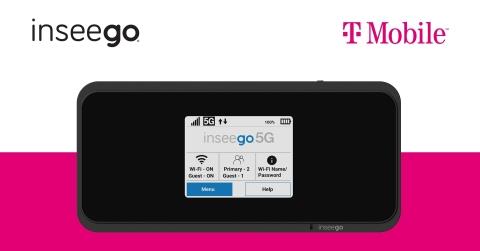}
	\caption{Inseego 5G MiFi® M2000}
	\label{fig:inseego}
\end{figure}

\begin{figure}[htbp]
	\centering
	\includegraphics[width=0.8\linewidth]{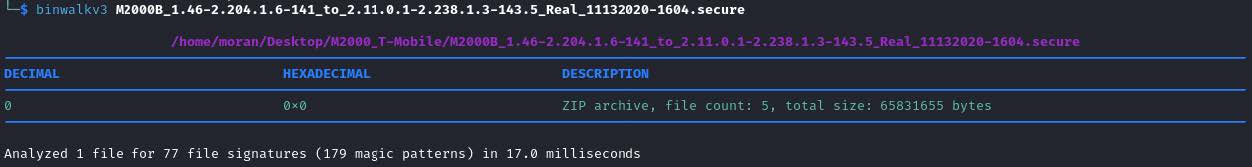}
	\caption{Firmware File Type and binwalk Analysis}
	\label{fig:binwalk-analysis}
\end{figure}

\begin{figure}[htbp]
	\centering
	\includegraphics[width=0.8\linewidth]{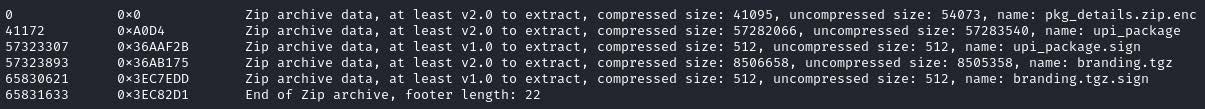}
	\caption{binwalk Analysis of Firmware Inside Binary File}
	\label{fig:binwalk-analysis2}
\end{figure}

\begin{figure}[htbp]
	\centering
	\includegraphics[width=0.8\linewidth]{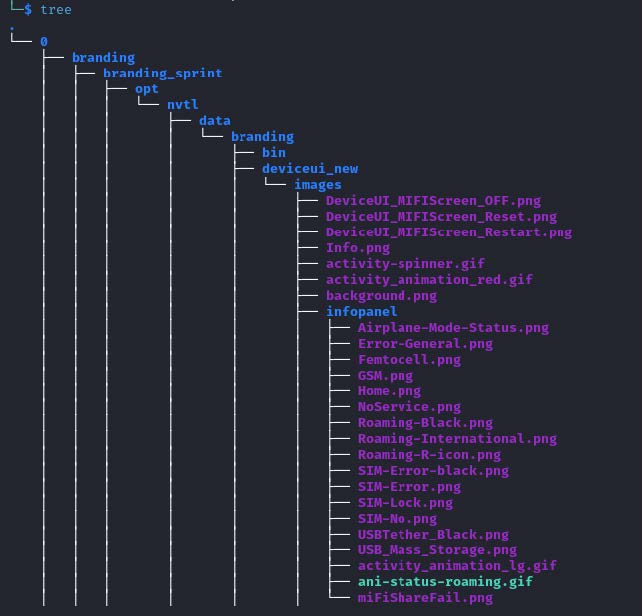}
	\caption{Part of the Firmware Inside Directory Tree}
	\label{fig:directory-tree}
\end{figure}

After we failed to access the real device and its firmware, we found a new direction to research T-Mobile security vulnerabilities. Our research revealed that T-Mobile has a partnership with Inseego, and T-Mobile has selected the Inseego 5G MiFi® M2000 as Its First 5G Mobile Hotspot, as shown in Figure \ref{fig:inseego}. In addition, we found the firmware of this device online\cite{m2000firmware}. After our preliminary analysis using binwalk, the firmware uses embedded custom Linux as the primary operating system. To prevent tampering and overflow, the firmware uses MD5 to check the integrity of the firmware and stores the final address of the system. There are many files in the firmware used to store the one-to-one correspondence between MD5, file paths and file names. Due to the large number of files, only the contents of two of them are listed here, as shown in Table \ref{tab:password-comparison}. In addition, some files in the firmware imply the existence of a signature mechanism for firmware integrity verification or authorisation purposes. In the firmware package inside mode, tar, zip, and xz are used for compression, as shown in Figure \ref{fig:binwalk-analysis}. The firmware suffix is ".secure" to confuse the firmware file type shown in Figure \ref{fig:binwalk-analysis2}. Additionally, part of the firmware is included in the directory tree, as shown in Figure \ref{fig:directory-tree}. Also, this device uses a Qualcomm CPU. 

\begin{table}[htbp]
	\centering
	\caption{Password Comparison Before and After Encryption}
	\label{tab:password-comparison}
	\begin{tabular}{|c|c|c|}
		\hline
		\textbf{Filename} & \textbf{MD5} & \textbf{Final Address} \\
		\hline
		\texttt{boot.old\_info} & f4454ec67c186a9532e0116888941969 & 7215104 \\
		\hline
		\texttt{boot.new\_info} & 7ba33035cf260c54f90243471efdd8d2 & 7227392 \\
		\hline
	\end{tabular}
\end{table}

\subsection{Cybersecurity Recommendations for Risk Mitigation}
\label{subsec:Cybersecurity Recommendations for Risk Mitigation}

\subsubsection{Continue Testing Beyond Perimeter}
\label{Continue Testing Beyond Perimeter}

Since we didn't have internal testing access to T-Mobile and due to the geographical distance, we only tested the devices exposed on the Internet. However, very often, threats come from within the enterprise, the enterprise's wireless network, and the physical equipment of the enterprise. We strongly recommend that privileges be used for more in-depth internal penetration testing during security audits to identify enterprise vulnerabilities, logical flaws, and misconfigurations. 

Within an enterprise, commercial spies or employees without cybersecurity awareness may exist. Such situations often pose a very significant threat to the enterprise. If an authorised employee is subject to social engineering or phishing attacks and the attacks succeed, it poses a considerable security threat to the enterprise. 

Enterprise wireless networks are typically found in business and office areas, often with visitor networks that lack strong passwords or have no password protection. If the isolation between the visitor network and the company's internal network is not well done, attackers are very likely to penetrate the enterprise's internal network through the visitor network. Therefore, it is necessary to strengthen the management of the visitor network, formulate standard norms, isolate each device in the visitor network, and prohibit devices from accessing the company's internal network devices. Secondly, implement network speed restrictions to prevent the visitor network from occupying excessive traffic. The internal wireless network of the company should also be appropriately set up and managed, and the wireless network scope should be planned through wireless network geological exploration as needed. Do not overly pursue coverage. Adjust everything according to the needs of the department and business, and try to prevent the internal wireless network signal from spreading outside the company. 

Many enterprises' physical devices may also be among the targets of attackers, such as the access control system, elevator system, and base station of the T-Mobile Company building. If T-Mobile uses M1 Card as its access control and elevator access rights systems, many security problems will arise. Although many enterprises do this, the encryption protocol and algorithm of the M1 Card were completely cracked as early as more than ten years ago. Therefore, non-fully encrypted M1 cards are very easy for attackers to copy within tens of seconds. Even fully encrypted cards can be cracked entirely and copied by attackers only twice. Therefore, a biometric system should be gradually adopted to replace the old M1 card access control system. The elevator system is the same. In addition, parking lots are also one of the scenarios where attackers can tamper with things. They may place information collection devices in parking lots, such as micro cameras, microphones, and wireless signal capture devices, to infiltrate enterprises. Therefore, it is also necessary to enhance the security of enterprise parking lots and conduct regular wireless signal scans of the parking lots to ensure that no unauthorised devices are placed there. 

In addition to strictly implementing the above plan, T-Mobile can hire some well-known ethical hacking auditing companies, such as Hak5 Company\footnote{Hak5 is a U.S.-based cybersecurity company founded in 2005, known for developing innovative penetration testing tools and producing award-winning infosec media. Their product lineup includes devices like the WiFi Pineapple, USB Rubber Ducky, and Bash Bunny—widely used by ethical hackers and red teams for wireless auditing, payload delivery, and covert access. Hak5 also fosters an inclusive hacker community and offers immersive training experiences tailored for cybersecurity professionals and enthusiasts\cite{hak5}.}, to conduct entity security penetration tests. They produce various ethical hacking teaching tools and create videos showing the security penetration of enterprise entities to raise awareness. The physical security penetration process may involve the use of actual social engineering attacks, lock-picking skills, wireless network audits, access control system penetration, etc. 

\subsubsection{Harden API and Endpoint Exposure}
\label{subsubsec:Harden API and Endpoint Exposure}
\begin{enumerate}[left=1em]
	\item \textbf{Enforce Token Validation:} When a user invokes an API request, the API gateway layer verifies whether the user's Access Token is valid, and a user with an invalid Token will not be able to invoke the API.
	
	\item \textbf{Use Role-Based Access Control (RBAC):} The API calls are divided and controlled according to the user's role. Usually, users can only call the basic API functions. They cannot access sensitive information through the API, which prevents unauthorised users from calling APIs other than those to which they have permission.
	
	\item \textbf{Optimise Token Life Cycle:} Each Access Token has a short validity period, after which it will be immediately destroyed and the corresponding user will be logged out.
	
	\item \textbf{Deploy Rate Limiting:} Configure the frequency limit for API invocation. When the same IP address or account frequently requests the same API within a short period of time, the system will temporarily block the permission of that IP address or account to invoke the API and record abnormal information in the log system.
	
	\item \textbf{Enhance Logging and Anomaly Detection:} Real-time collection and storage of access logs for key resources (such as API gateways and authentication services). Use the SIEM system to monitor and analyse access logs continuously, automatically detect pre-configured abnormal behaviours, and generate alerts.
	
	\item \textbf{Amazon GuardDuty:} Amazon GuardDuty is a managed threat detection service provided by AWS, capable of achieving the functions of Deploy Rate Limiting, Enhanced Logging, and Anomaly Detection. Amazon GuardDuty automatically integrates and analyses VPC Flow Logs, CloudTrail\footnote{AWS CloudTrail is a governance and auditing service that records API calls and user activity across AWS services. It provides detailed event logs for actions taken via the AWS Management Console, CLI, SDKs, and APIs, enabling security analysis, operational troubleshooting, and compliance reporting. CloudTrail supports multi-account and multi-region logging, and integrates with services like Amazon S3, CloudWatch, and CloudTrail Lake for long-term storage and advanced analytics\cite{aws_cloudtrail_doc}.} management events, and data events, as well as DNS query logs. It uses machine learning models to analyse standard behaviour patterns in accounts and identify deviations, such as an account that frequently logs in from the same IP suddenly logging in from different IPs, which is considered abnormal behaviour. It uses the malicious IP/domain database provided by AWS to identify malicious addresses. It employs behavioural analysis models to detect unusual behaviours. For example, if a user who rarely accesses S3 buckets suddenly downloads tens of thousands of files consecutively, this is also considered abnormal behaviour. Based on the three methods mentioned above, Amazon GuardDuty marks the severity level of detected anomalies, identifies threat types, and can optionally automatically push alerts to Amazon CloudWatch or AWS SNS and AWS Lambda for generating alerts, email, SMS notifications, and other security operations\cite{guarddutyeventbridge} \cite{guarddutycloudwatch} \cite{whatisguardduty}. 
\end{enumerate}

\subsubsection{IoT Security Best Practices}
\label{IoT Security Best Practices}

For IoT devices, it is recommended to use OpenSSL for encryption and firmware integrity verification. Do not use other compression algorithms or obscure file extensions to protect the firmware. Some other enterprises use another encryption method to encrypt the firmware. Hikvision\footnote{Hikvision is a Chinese multinational technology company specialising in video surveillance products and solutions. As one of the world’s largest manufacturers of security cameras and AIoT systems, Hikvision offers a wide range of devices, including IP cameras, NVRs, access control systems, and intelligent analytics platforms.} uses firmware with the ".dav" extension on its camera or video recorder devices. Then they use an encryption algorithm like Caesar encryption to encrypt the content of the firmware file. This indeed avoids the initial scan of binwalk. However, it is still very easy to decrypt if an expert manually analyses it. Therefore, the following OpenSSL script is recommended for firmware encryption. The encryption script is shown in Listing \ref{lst:encrypt-firmware}. The decryption script is placed in the IoT device as shown in Listing \ref{lst:decrypt-firmware}. Please note that when using this script, you should set the password as a system variable or use a hash-encrypted password when the device exits. Do not store the plaintext of the password in the script. Otherwise, the password will be leaked if someone removes the flash chip to read its content. Moreover, different models and versions of devices are recommended to use other strong password generators to generate passwords for them. 

\begin{lstlisting}[language=bash, caption={Encrypt Firmware Bash}, label={lst:encrypt-firmware}]
	#!/bin/bash
	
	INPUT_FW="firmware.bin"
	ENCRYPTED_FW="firmware.enc"
	OUTPUT_FILE="firmware.secure.bin"
	FW_PASS="${FW_PASS:StrongPassword123}"
	
	openssl enc -aes-256-cbc -salt -pbkdf2 -iter 100000 \
	-in "$INPUT_FW" -out "$ENCRYPTED_FW" -pass pass:"$FW_PASS"
	
	HASH=$(sha256sum "$ENCRYPTED_FW" | awk '{print $1}')
	printf "%s" "$HASH" > firmware.hash
	
	cat "$ENCRYPTED_FW" firmware.hash > "$OUTPUT_FILE"
	
	echo "[+] Encryption Completed: $OUTPUT_FILE"
\end{lstlisting}

\begin{lstlisting}[language=bash, caption={Decrypt Firmware Bash}, label={lst:decrypt-firmware}]
	#!/bin/bash
	
	IN_FILE="/tmp/firmware.secure.bin"
	TMP_ENC="/tmp/firmware.enc"
	TMP_DEC="/tmp/firmware.bin"
	TMP_HASH="/tmp/firmware.hash"
	HASH_LEN=64
	FW_PASS="${FW_PASS:StrongPassword123}"
	
	BIN_SIZE=$(stat -c%s "$IN_FILE")
	ENC_SIZE=$((BIN_SIZE - HASH_LEN))
	
	dd if="$IN_FILE" bs=1 count=$ENC_SIZE of="$TMP_ENC" 2>/dev/null
	tail -c $HASH_LEN "$IN_FILE" > "$TMP_HASH"
	
	ACTUAL_HASH=$(sha256sum "$TMP_ENC" | awk '{print $1}')
	EXPECTED_HASH=$(cat "$TMP_HASH")
	
	echo "$ACTUAL_HASH"
	echo "$EXPECTED_HASH"
	
	if [ "$ACTUAL_HASH" != "$EXPECTED_HASH" ]; then
	echo "[!] Verification failed"
	exit 1
	fi
	
	echo "[+] Verification successful, starting decryption..."
	
	openssl enc -d -aes-256-cbc -pbkdf2 -iter 100000 \
	-in "$TMP_ENC" -out "$TMP_DEC" -pass pass:"$FW_PASS"
	
	echo "[+] Decryption completed, ready to upgrade"
\end{lstlisting}

\begin{figure}[htbp]
	\centering
	\includegraphics[width=0.8\linewidth]{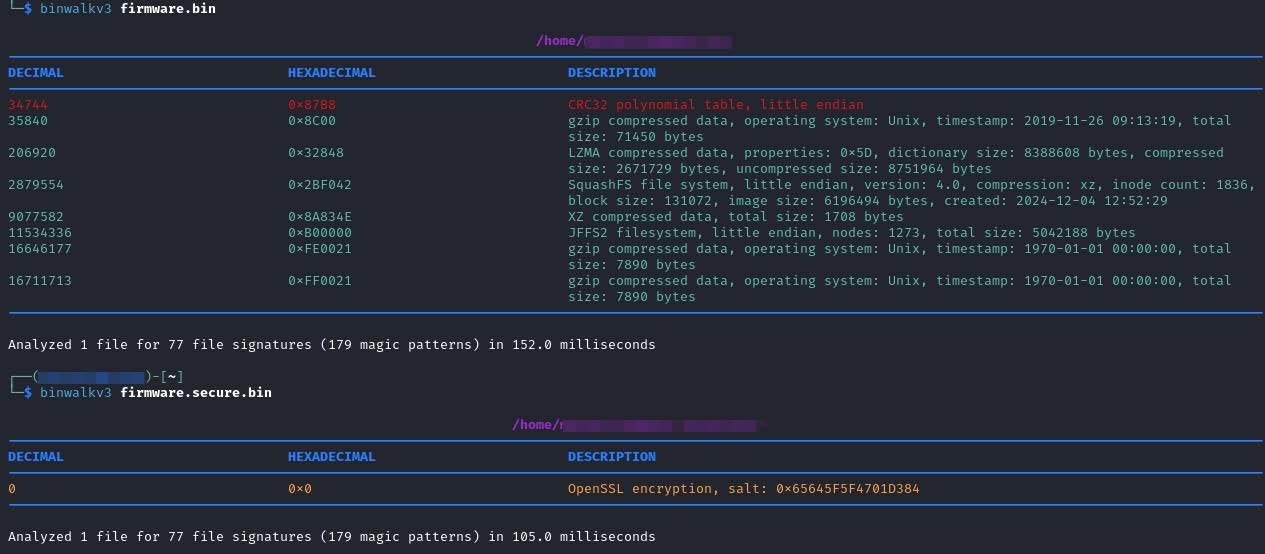}
	\caption{Detect Firmware Structure Using binwalkv3}
	\label{fig:firmware-structure}
\end{figure}

Figure \ref{fig:firmware-structure} shows the comparison result between an unencrypted firmware and the firmware encrypted by the above script using binwalkv3. It can be seen from this that binwalk can no longer detect the structure and content within it, only detecting OpenSSL encryption. 

Additionally, corresponding security measures and device verification mechanisms, such as IPS/WAF, should also be deployed on the websites or file servers for IoT device upgrades. This is to prevent someone from disassembling the flash chip and exporting the firmware after the entity obtains the device, obtains the address of the firmware update, and launches an attack on the firmware update server. Once the attack is successful, it will likely cause a wide range of impacts. Based on T-Mobile's past experiences, they often overlook this point, assuming that as long as the server address is not exposed, the server is secure. 

\subsubsection{Secure Web Application Development Practices}
\label{subsubs:Secure Web Application Development Practices}

From the above tests on the Web, many minor issues can be identified, such as missing flag Settings, failure to enforce secure transmission, and tokens being within the URL. Overall, no severe problems were found on the main website facing customers, but to prevent these problems from causing major loopholes in the future, further measures are necessary. T-Mobile should undergo a comprehensive upgrade. The reason for these vulnerabilities might be that the website is maintained by different programmers and personnel. Some of the code on the website is rather outdated. A plan needs to be made to upgrade it while ensuring the availability of customer service gradually. 

In addition, when we scanned the 'T-Mobile Careers' website, we discovered a vulnerability in the plaintext transmission of account passwords. Although this website is not the main customer service website of T-Mobile, its traffic is relatively small. Also, the website applies TLS to encrypt traffic. However, suppose a zero-day vulnerability like Heartbleed emerges that can decrypt TLS traffic or bypass the encryption in some way. In that case, the vulnerability of this website is very likely to affect the entire company, just as Snowden was able to expose the Prism scandal\footnote{The PRISM scandal refers to the 2013 revelations by whistleblower Edward Snowden about a classified surveillance program operated by the U.S. National Security Agency (NSA). Under PRISM, the NSA collected internet communications from major tech companies like Google, Microsoft, and Facebook through court-approved requests under Section 702 of the FISA Amendments Act. The disclosures sparked global debate over privacy, government overreach, and the legality of mass data collection, leading to legal reforms and heightened scrutiny of intelligence practices\cite{wikipediaPRISM}.}. For a large company like T-Mobile, effectively managing its branch companies and other service devices is crucial. The solution to this problem is to use the back-end JavaScript code to encrypt the account password submitted by the user. The encryption method used is AES, and the generation method of Nonce should be set to increase the cracking cost for attackers and make the encryption more secure. The existence of a suitable Nonce can ensure that even when users submit the same password, the ciphertext will be completely different under the avalanche effect. Finally, SHA384 was used to hash the encrypted password. To facilitate the understanding of the above process, it is more intuitive to implement it with Python. The Python code is shown in Listing \ref{lst:python-password-update}. 

\begin{lstlisting}[language=Python, caption={Python Password Update Functions}, label={lst:python-password-update}]
	def nonce_create(self):
	type_val = 0
	device_id = 'b7:e0:6a:f2:8e:c3'
	time_val = int(time.time())
	random_val = random.randint(0, 10000)
	return f"{type_val}_{device_id}_{time_val}_{random_val}"
	
	def new_pwd(self, pwd, new_pwd):
	key = hashlib.sha384((pwd + self.key).encode()).hexdigest()[:32]
	key = bytes.fromhex(key)
	password = hashlib.sha384((new_pwd + self.key).encode()).hexdigest()
	iv = bytes.fromhex(self.iv)
	return password
\end{lstlisting}

Listing \ref{lst:python-password-update} can work properly after defining a key and an IV. The Two Nonce and hashed encrypted password examples are shown in Table \ref{tab:password-comparison2}. 

\begin{table}[htbp]
	\centering
	\caption{Password Comparison Before and After Encryption}
	\label{tab:password-comparison2}
	\begin{tabular}{|c|c|c|}
		\hline
		\textbf{Plaintext Password} & \textbf{Nonce} & \textbf{Encrypted and Hashed Password} \\
		\hline
		\begin{tabular}[c]{@{}c@{}}newpassword\\456\end{tabular} &
		\begin{tabular}[c]{@{}c@{}}0\_b7:e0:6a:f2:8e:c3\_1747561236\_\\2745\end{tabular} &
		\begin{tabular}[c]{@{}c@{}}83bd0dd9e493830e7d43a809110a3302838ecd8de7716b362\\b4cdd6192d00f9600d35780cbe0aac539e7eb9775386389\end{tabular} \\
		\hline
		\begin{tabular}[c]{@{}c@{}}newpassword\\456\end{tabular} &
		\begin{tabular}[c]{@{}c@{}}0\_b7:e0:6a:f2:8e:c3\_1747561263\_\\2715\end{tabular} &
		\begin{tabular}[c]{@{}c@{}}42aa51a0af90f29116b1587f10e655b8feefd856c2e322ec104\\2e74a38be85d8349cfac2f2535becae3666ae4d617d01\end{tabular} \\
		\hline
	\end{tabular}
\end{table}

\subsubsection{Device and Remote Access Security Controls}
\label{subsubsec:Device and Remote Access Security Controls}
How to manage devices safely via the Internet or remotely has always been a significant problem T-Mobile faces. In the above test, we discovered an unencrypted and unlimited-attempt VNC interface of T-Mobile. Subsequently, we attempted brute-force cracking of passwords. Although the cracking was unsuccessful, it can still indicate that T-Mobile has many improper configuration problems in this aspect. Notably, this VNC misconfiguration is identical to the SSH configuration issue that led to the 2021 data breach. Obviously, T-Mobile does not have a truly effective policy for safely managing devices remotely. This problem might be due to the behaviour of individual employees or vulnerabilities not discovered by the company's security audit. The solution to this problem is that the company's internal security training program must be strictly implemented, and it is necessary to ensure that every authorised employee has a strong awareness of cybersecurity, as we said in Section \ref{sec:Comprehensive Infrastructure and Security Enhancement Strategy}. Secondly, when conducting security audits, it is necessary to refer to the vulnerability scanning data from third-party public engines, such as Shodan. T-Mobile can purchase Shodan's enterprise services, and after each scan by Shodan detects a problem, security engineers will be arranged to troubleshoot each issue one by one manually. Further, a set of security automation deployment strategies and corresponding procedures should be formulated to ensure that this script is used for configuration when the company has new equipment to avoid problems and vulnerabilities caused by manual configuration. Finally, a professional honeypot system is deployed on the server to record the attack behaviours of hackers, enrich the evidence after future network attacks, observe the attack methods of hackers, and formulate corresponding security strategies. 

\subsubsection{Maintain Regular Third-Party Penetration Tests}
\label{subsubsec:Maintain Regular Third-Party Penetration Tests}

Choose a third-party penetration test with national or internationally recognised security testing qualifications in the following areas in Table \ref{tab:test-items}

\begin{table}[htbp]
	\centering
	\caption{Test items}
	\label{tab:test-items}
	\begin{tabular}{|c|c|c|}
		\hline
		\textbf{Project type} & \textbf{Recommendation Frequency} & \textbf{Test Mode} \\
		\hline
		\begin{tabular}[c]{@{}c@{}}Web applications/API\\ for the public network\end{tabular} &
		Every six months &
		Black box and grey box \\
		\hline
		\begin{tabular}[c]{@{}c@{}}IoT firmware and\\ gateway interface\end{tabular} &
		Every year &
		Firmware reverse engineering and API security testing \\
		\hline
		\begin{tabular}[c]{@{}c@{}}Internal networks\\ and systems\end{tabular} &
		Every year &
		White box testing and social engineering simulation \\
		\hline
		\begin{tabular}[c]{@{}c@{}}Key business\\ processes\end{tabular} &
		Every three months &
		Scenario simulation test, simulated attack path \\
		\hline
	\end{tabular}
\end{table}

\subsubsection{Prioritisation Plan}
\label{subsubsec:Prioritisation Plan}
Based on the above test results, we have formulated the priority solution schedule, which is shown in Table \ref{tab:prioritisation-plan}. T-Mobile can optimise according to the following table. 

\renewcommand{\arraystretch}{1.1}

\begin{table}[H]
	\centering
	\caption{Prioritisation Plan}
	\label{tab:prioritisation-plan}
	\begin{tabular}{|
			>{\centering\arraybackslash}m{5.5cm}|
			>{\centering\arraybackslash}m{7cm}|
			>{\centering\arraybackslash}m{2cm}|}
		\hline
		\textbf{Vulnerability} & \textbf{Reason} & \textbf{Priority} \\
		\hline
		Career Website transmits the login credentials in plaintext &
		The username and password are submitted in plain text, allowing an attacker to steal sensitive information through a man-in-the-middle attack. &
		High \\
		\hline
		Shodan Exposure of devices with high-risk vulnerabilities, such as SMBv3 RCE &
		Remote code execution exists on some devices. &
		High \\
		\hline
		VNC service exposure + no encryption enabled + no blasting protection mechanism &
		It can be cracked by brute force and easily become an entrance to remote control. &
		Medium \\
		\hline
		The historical API/URL has potential injection points &
		There is a suspected SQL parameter structure in the old interface that the current WAF detection may bypass. &
		Medium \\
		\hline
		Nested third-party pages have medium risk issues &
		Embedded or integrated from third-party services, there is a lack of a unified vulnerability response process. &
		Medium \\
		\hline
		The API returns too much information &
		It may be used for information gathering. &
		Low \\
		\hline
		\texttt{gobuster/dirsearch} scan did not find a valid entry &
		No direct risk, but continuous monitoring is required. &
		Low \\
		\hline
	\end{tabular}
\end{table}

\section{Conclusions}
\label{conclusions}
Through a combined strategy of breach forensics and systemic vulnerability testing, this study reveals how latent architectural flaws and inconsistent policy enforcement remain high-risk vectors in T-Mobile’s infrastructure, even after remediation efforts. From unsecured remote interfaces and plaintext credential submission to firmware supply chain exposure, the results of our ethical hacking audit confirm that true resilience requires defence-in-depth at every layer. 

The proposed countermeasures—including AWS-integrated API gateways, endpoint rate control, encrypted firmware lifecycle management, and rigorous access auditing—demonstrate that enterprise-scale protection is both technically feasible and economically justified. Importantly, our audit process simulates real-world attacker behaviour across web, device, and cloud boundaries, providing a template for other telecommunications firms confronting post-breach security recovery. 

Future research will extend to dynamic anomaly detection using AI models, firmware integrity proofing using zero-knowledge systems, and evaluating cross-provider security interdependencies in global telecom ecosystems. 

\bibliography{main}

\end{document}